\documentclass{aa}
\usepackage{graphicx}
\usepackage{amsfonts,amssymb}
\usepackage{txfonts}

\def\gsim{\lower.4ex\hbox{$\;\buildrel >\over{\scriptstyle\sim}\;$}} 
\def\lsim{\lower.4ex\hbox{$\;\buildrel <\over{\scriptstyle\sim}\;$}} 

\def\[{\begin{eqnarray}}
\def\]{\end{eqnarray}}

\def\q{\qquad}
\def\qq{\qquad\qquad}

\newcommand{\Rm}{\mathrm{Rm}}

\newcommand{\Rin}{R_\mathrm{in}}
\newcommand{\Rout}{R_\mathrm{out}}

\newcommand{\etah}{\hat{\eta}}

\newcommand{\cmnt}[1]{}
\newcommand{\comm}[1]{}
\newcommand{\ignore}[1]{}

\def\gsim{\lower.4ex\hbox{$\;\buildrel >\over{\scriptstyle\sim}\;$}} 
\def\lsim{\lower.4ex\hbox{$\;\buildrel <\over{\scriptstyle\sim}\;$}} 
\def\qq{\qquad\qquad}                      
\def\q{\qquad}
\def\beg{\begin{eqnarray}}
\def\ende{\end{eqnarray}}
\renewcommand{\vec}[1]{\mbox{\boldmath $#1$}}

\newcommand{\Om}{{\it \Omega}}

\begin{document}

 \title{Stratorotational instability in MHD Taylor-Couette flows}

 \author{G. R\"udiger\inst{1}
    \and D.A. Shalybkov\inst{2}}
 
 \institute{
 Astrophysikalisches Institut Potsdam, An der Sternwarte 16, 
 D-14482 Potsdam, Germany
\and
A.F. Ioffe Institute for Physics and Technology, 194021, St. Petersburg, Russia}

\date{\today; accepted}
 \abstract{} 
 {The stability of dissipative Taylor-Couette flows with an axial stable
density stratification and a prescribed  azimuthal magnetic field is considered. } 
{ Global nonaxisymmetric solutions of the linearized MHD equations with toroidal magnetic field, axial density stratification and differential rotation are found for both insulating and conducting cylinder walls.} 
{Flat rotation laws such as the quasi-Kepler  law are  
unstable against the nonaxisymmetric stratorotational instability (SRI). The influence of a
current-free toroidal magnetic field   depends on the
magnetic Prandtl number Pm: SRI is supported  by $\rm Pm > 1$ and it is suppressed 
by  $\rm Pm \lsim 1$.  For too flat rotation laws a smooth transition exists to the instability  which the toroidal magnetic field produces in combination with the differential rotation. This nonaxisymmetric azimuthal magnetorotational instability (AMRI) has been computed   under the presence of    an axial density gradient.

If the magnetic field between the cylinders is not current-free then
also the  Tayler instability  occurs and the transition from the
hydrodynamic SRI to the magnetic Tayler instability  proves to be rather
complex. Most spectacular is the `ballooning' of the stability domain by the density stratification: 
already a rather small rotation stabilizes  magnetic fields against the 
 Tayler instability.

An  azimuthal component of the resulting  electromotive force    only exists for density-stratified flows. The related  alpha-effect  for
magnetic SRI of  Kepler rotation appears   to be positive for negative $d\rho/dz <0$.
}  
{}

 
\authorrunning{G. R\"udiger \& D.A. Shalybkov}
 \titlerunning{Stratorotational instability in MHD Taylor-Couette flow}
 \maketitle

 \section{Motivation}
 This work is motivated
by  theoretical and experimental progresses in studies of  the stratorotational instability (SRI) and the magnetorotational instability (MRI) in MHD Taylor-Couette experiments. It has been shown  theoretically  (Molemaker,   McWilliams  \& Yavneh 2001; Yavneh,   McWilliams \&  Molemaker 2001; Dubrulle et al. 2005; Shalybkov \&  R\"udiger 2005; Umurhan 2006) and in the laboratory (Le Bars \&  Le Gal 2007)
that a combination of  a centrifugal-stable nonuniform rotation law  and a stable axial density stratification leads to the  so-called stratorotational instability (SRI) in the Taylor-Couette flow.
This instability exists only for nonaxisymmetric disturbances.
On the other hand, there are also   nonaxisymmetric instabilities for a combination of Rayleigh-stable
rotation laws  and  azimuthal magnetic fields
(Pitts \& Tayler 1985).
The question is whether the combination of density stratification, differential rotation and toroidal fields acts stabilizing or destabilizing or whether even new instabilities arise.

Such a  combination of axial density stratification, stable rotation law and strong toroidal magnetic field is the typical constellation in accretion disks  (Ogilvie \&  Pringle 1996; Curry \& Pudritz 1996; Papaloizou \& Terquem 1997). There the rotation is Keplerian  with $\Om\propto R^{-3/2}$ and the toroidal field is generated from weak large-scale poloidal fields by  induction due to the  differential rotation. The standard case is that the resulting toroidal field strongly exceeds the amplitude of the original poloidal field if the magnetic Reynolds number of the differential rotation $\Rm =\Om R^2/\eta$ is much larger than unity. If 
this is true then the standard MRI, i.e. the influence of the poloidal field on the stability of the differential rotation, would be of minor importance\footnote{the same is true for calculations about  the stability of strong poloidal fields without or with rotation}. The question is whether the toroidal magnetic field can reach such high amplitudes or whether it becomes unstable already for much smaller values. It is known that  toroidal fields with  strong electric currents become unstable (Tayler instability, `TI') but it is also known that in combination with differential rotation also toroidal fields become unstable which are current-free in the considered domain (azimuthal MRI, `AMRI', see  R\"udiger et al. 2007a). In the latter 
case, with $B_\phi\propto R^{-1}$, the questions are whether the density stratification destabilizes AMRI and/or whether the toroidal field stabilizes the SRI too strongly so that its real existence becomes basically suppressed.

We shall consider the interaction of the differential rotation  with  both an axial density
stratification and a toroidal  magnetic field in the simplified Taylor-Couette geometry. The density stratification is always supposed to be 
stable but the magnetic field and the rotation law between the cylinders can be both  stable or
unstable. If, in particular, the toroidal field is Tayler unstable then the interaction of differential rotation, density stratification and magnetic field becomes highly complex. We find stabilization and destabilization in strong dependence on the magnetic Prandtl number. Again, in experiments, for small magnetic Prandtl number,  the flows are predicted to be  stabilized. For galaxies and protoneutron stars (PNS) with their high magnetic Prandtl numbers  
we find the opposite: the magnetic influence supports  the SRI leading  to even smaller Reynolds numbers than in hydrodynamics.

\section{The Taylor-Couette geometry}
\begin{figure}
    \center \includegraphics[width=6.0cm]{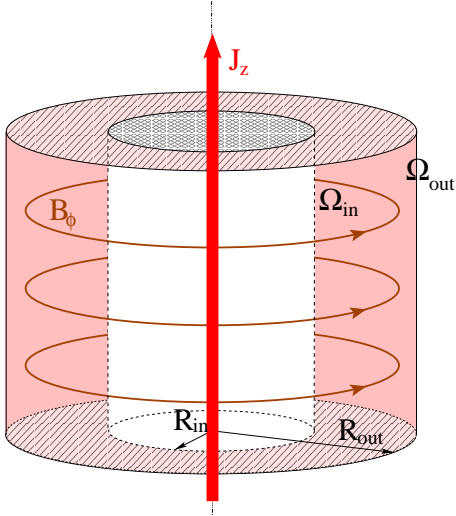}
    \caption{The geometry of the problem -- two concentric cylinders with radii
	     $\Rin$  and  $\Rout$  rotating with $\Om_{\rm in}$ and
	     $\Om_{\rm out}$. $B_\phi$ is the azimuthal magnetic
	     field which, generally, is produced by  both an axial current inside the inner
	     cylinder and an axial current through the fluid. }
    \label{geom}
 \end{figure}

A Taylor-Couette container is considered which confines  a toroidal magnetic field with
given amplitudes at the cylinder walls which rotate with different rotation rates $\Om$ 
(see Fig. \ref{geom}). 

The fluid  between the cylinders is assumed to be incompressible and
 dissipative with the kinematic viscosity $\nu$ and the magnetic diffusivity
 $\eta$. Derived from  the conservation of angular momentum the rotation law
 $\Om(R)$ in the fluid is
\beg
    \Om=a+\frac{b}{R^2}
\label{basicOm} 
\ende
 with
 \beg
  a=\frac{\mu_\Om-\etah^2}{1-\etah^2} \Om_{\rm in}, \ \ \ \ \ \ \ \ \ \ 
  b=\frac{1-\mu_\Om}{1-\etah^2}\Rin^2 \Om_{\rm in},
 \ende
 and
\beg
\hat\eta=\frac{R_{\rm{in}}}{R_{\rm{out}}}, \ \ \ \ \ \ \ \ \ \ \ \ \ \ \ \ \ \ \ \ \ \ 
\mu_\Om=\frac{\Om_{\rm{out}}}{\Om_{\rm{in}}},
\label{mu}
\ende
where  cylindric coordinates ($R$,$\phi$,$z$) are used and
$\Om_{\rm in}$ and $\Om_{\rm out}$ are the imposed rotation rates at
the inner and outer cylinders of radii $R_{\rm in}$ and $R_{\rm out}$.

Similarly, the magnetic profiles are restricted by the azimuthal component of the induction equation to
\beg
B_\phi=A R+\frac{B}{R},
\label{basicB}
\ende
where the first term corresponds to a uniform axial
electric current at radius $R$ and the second term is current free for $R>0$. 
 In analogy with
$\mu_\Omega$ it is useful to define the quantity
\beg
\mu_B=\frac{B_{\rm{out}}}{B_{\rm{in}}}
  =\frac{A R_{\rm out}+B R_{\rm out}^{-1}}
        {A R_{\rm in }+B R_{\rm in }^{-1}},
\ende
measuring the variation in $B_\phi$ across the gap between the cylinders.
The coefficients $A$ and $B$ are
\beg
A=\frac{B_{\rm{in}}}{R _{\rm{in}}}\frac{\hat \eta
(\mu_B - \hat \eta)}{1- \hat \eta^2},  \ \ \ \ \ \ \ \ \ \ \ \ \ \ \ \ \ \ \ \ 
B=B_{\rm{in}}R _{\rm{in}}\frac{1-\mu_B \hat\eta}
{1-\hat \eta^2}.
\label{ab}
\ende
The MHD equations for incompressible stratified fluids  are
\begin{eqnarray}
\lefteqn{\frac{\partial \vec{u}}{\partial t} + (\vec{u} \nabla)\vec{u} =
-\frac{1}{\rho} \nabla P +\vec{g} + \nu \Delta \vec{u}+
 \frac{1}{\mu_0}{\rm curl}\vec{B} \times \vec{B},}
\nonumber \\
\lefteqn{\frac{\partial \vec{B}}{\partial t}= {\rm curl} (\vec{u} \times \vec{B})
+ \eta \Delta\vec{B},}
\nonumber \\
\lefteqn{\frac{\partial \rho}{\partial t}+(\vec{u}\nabla)\rho=0,}
\nonumber \\
\lefteqn{{\rm and}}\nonumber\\
\lefteqn{{\rm div}\ \vec{u} = 0, \ \ \ \ \ \ \ \ \ \  {\rm div}\ \vec{B} = 0,}
\label{mhd}
\end{eqnarray}
where $\vec{u}$ is the velocity, $\vec{B}$ is the magnetic field, $P$ is the pressure,
$g$ is the gravitational acceleration (supposed as vertical and constant),
$\nu$ is the kinematic viscosity, $\eta$ is the magnetic diffusivity and
$\mu_0$  the magnetic constant.

In the presence of a vertical density gradient ($\rho=\rho(z)$) it has been shown for $\vec{B}=0$, that the system (\ref{mhd}) allows the angular velocity profile
(\ref{basicOm}) only in the limit of slow rotation and small stratification
\beg
\left|\frac{R\Om^2}{g}\right| \ll 1, \q \left|\frac{d{\rm log}(\rho)}{d{\rm log}(z)}\right| \ll 1.
\label{gcond}
\ende
It easy to show that the same is true for $B\neq 0$.
So, we are interested in the stability of the basic state
\begin{eqnarray}
\lefteqn{\vec{U}=(0,R\Om(R),0), \ \ \ \ \ \ \ \ \ \ \ \ \ \ \ \ \ \ \ \ \ \ \ \ \ \ \ \vec{B}=(0,B_\phi(R),0),}
\nonumber \\
\lefteqn{P=P_0(R)+P_1(R,z), \ \ \ \ \ \ \ \ \ \ \ \ \ \ \ \ \ \ \ \ \ \ \rho=\rho_0+\rho_1(z),}
\label{basicst}
\end{eqnarray}
where $\Om$ is given by (\ref{basicOm}), $B_\phi$  by (\ref{basicB}),
$\rho_0$ is the uniform reference density, $P$ is the total pressure including the magnetic part with
$ |P_1/P_0| \ll 1$ and $|\rho_1/\rho_0| \ll 1$.

The linear stability problem is considered for the perturbed state of 
$\vec{U}+\vec{u}$, $\vec{B}+\vec{b}$, $\rho_0+\rho_1+\rho'$, $P_0+P_1+P'$.
Using the conditions (\ref{gcond}) the 
linearized system (\ref{mhd}) takes the Boussinesq form with the coefficients depending
only on the radial coordinate, and a normal mode expansion of the solution
$F=F(R){\rm exp}({\rm i}(m\phi+kz+\omega t))$ can be used, where $F$ represents any of the disturbed
quantities. 

Finally after a normalization we  find
\begin{eqnarray}
\lefteqn{\frac{ {\rm d}^2 u_R}{ {\rm d} R^2}
+\frac{1}{R}\frac{{\rm d} u_R}{{\rm d} R}
-\frac{u_R}{R^2} -\left(k^2+\frac{m^2}{R^2}\right) u_R
-2{\textrm{i}}\frac{m}{R}u_\phi-}
\nonumber \\
&& \qq -{\textrm{i Re}}(\omega+m\Om) u_R
+2{\textrm{Re}}\Om u_\phi
-\frac{ {\rm d} P'}{{\rm d} R}
\nonumber \\
&& \qq +{\textrm{i}}\frac{m}{R}{\textrm{Ha}}^2 B_\phi b_R
-2{\textrm{Ha}}^2 \frac{B_\phi}{R} b_\phi=0,
\nonumber \\
\lefteqn{\frac{ {\rm d}^2 u_\phi}{{\rm d} R^2}
+\frac{1}{R}\frac{{\rm d} u_\phi}{{\rm d} R}
-\frac{u_\phi}{R^2} -\left(k^2+\frac{m^2}{R^2}\right) u_\phi
+2{\textrm{i}}\frac{m}{R}u_R-}
\nonumber \\
&& \qq -{\textrm{i Re}}(\omega+m\Om) u_\phi
-{\textrm{i}}\frac{m}{R}P'
-\frac{{\textrm{Re}}}{R}\frac{{\rm d}}{{\rm d} R}(R^2 \Om) u_R
\nonumber \\
&& \qq +\frac{{\textrm{Ha}}^2}{R}\frac{{\rm d}}{{\rm d}R}\left(B_\phi R \right) b_R
+{\textrm{i}}\frac{m}{R}{\textrm{Ha}}^2 B_\phi b_\phi=0,
\nonumber \\
\lefteqn{\frac{ {\rm d}^2 u_z}{{\rm d} R^2}
+\frac{1}{R}\frac{{\rm d} u_z}{{\rm d} R}
-\left(k^2+\frac{m^2}{R^2}\right) u_z
-{\textrm{i Re}}(\omega+m\Om) u_z-}
\nonumber \\
&& \qq-{\textrm{i}}\,kP'-{\textrm{Re}}\,\rho'
+{\textrm{i}}\frac{m}{R}{\textrm{Ha}}^2 B_\phi b_z=0,
\nonumber \\
\lefteqn{{\textrm{i}}(\omega+m\Om)\rho'
-N^2u_z=0,}
\nonumber \\
\lefteqn{\frac{{\rm d} u_R}{{\rm d} R}+\frac{u_R}{R}
+{\textrm{i}}\frac{m}{R}u_\phi+{\textrm{i}}ku_z=0,}
\nonumber \\
\lefteqn{\frac{{\rm d}^2 b_R}{{\rm d} R^2}
+\frac{1}{R}\frac{{\rm d} b_R}{{\rm d} R}
-\frac{b_R}{R^2} -\left(k^2+\frac{m^2}{R^2}\right) b_R
-2{\textrm{i}}\frac{m}{R^2}b_\phi-}
\nonumber \\
&& \qq -{\textrm{i Pm Re}}(\omega+m\Om) b_R
+{\rm i}\frac{m}{R}B_\phi u_R=0,
\nonumber \\
\lefteqn{\frac{ {\rm d}^2 b_\phi}{{\rm d} R^2}
+\frac{1}{R}\frac{{\rm d} b_\phi}{{\rm d} R}
-\frac{b_\phi}{R^2} -\left(k^2+\frac{m^2}{R^2}\right) b_\phi
+2{\textrm{i}}\frac{m}{R^2}b_R-}
\nonumber \\
&&\qq -{\textrm{i Pm Re}}(\omega+m\Om) b_\phi
+{\rm Pm Re} R\frac{ d \Om}{{\rm d} R}b_R-
\nonumber \\
&&\qq -R\frac{{\rm d}}{{\rm d} R}\left(\frac{B_\phi}{R}\right) u_R
+{\rm i}\frac{m}{R}B_\phi u_\phi=0,
\nonumber \\
\lefteqn{\frac{{\rm d}^2 b_z}{{\rm d} R^2}
+\frac{1}{R}\frac{{\rm d} b_z}{{\rm d} R}
-\left(k^2+\frac{m^2}{R^2}\right) b_z-}
\nonumber \\
&&\qq -{\textrm{i Pm Re}}(\omega+m\Om) b_z
+{\rm i}\frac{m}{R}B_\phi u_z=0,
\label{sysf}
\end{eqnarray}
where the same symbols are used for the normalized quantities except $P'$ which denotes
$P'/\rho_0$ and redefining $\rho'$ as PmRe $g\rho'/\rho_0$. 
The dimensionless numbers of the problem are the magnetic
Prandtl number Pm, the Hartmann number Ha, the Reynolds number Re 
\begin{eqnarray}
\lefteqn{{\rm Pm}=\frac{\nu}{\eta}, \ \ \ \ \ \ \ \ \ \ \ \ \; {\rm Ha}=\frac{B_{\rm in}R_0}{\sqrt{\mu_0\rho\nu\eta}},
\ \ \ \ \ \ \ \ \ \ \ \ \ \ \ \ \ \  {\rm Re}=\frac{\Om_{\rm in}R_0^2}{\nu},}
\label{numbers}
\end{eqnarray}
 and the 
buoyancy frequency $N$ after
\beg
N^2=-\frac{g}{\rho_0}\frac{{\rm d} \rho_1}{{\rm d} z}.
\label{N2}
\ende
We used $R_0=(R_{\rm in}(R_{\rm out}-R_{\rm in}))^{1/2}$ as the unit of length,  $\eta/R_0$ as the
unit of the perturbation velocity, $B_{\rm in}$ as the magnetic field unit,
$\Om_{\rm in}$ as the unit of $\omega$, $N$, and $\Om$, $R_0\Om^2$ as the unit of
$g$, $\rho_0$ as the density unit and $\nu\eta/R_0^2$ as the unit of the redefined $P'$.
 It is convenient to describe the influences of the density stratification and the 
basic rotation by the Froude number
\beg
{\rm Fr}= \frac{\Om_{\rm in}}{N}.
\label{Fr}
\ende 

A detailed description of the  numerical methods  has 
been given in  earlier papers (see e.g. R\"udiger et al. 2007b) and will not be reproduced here.
Always no-slip boundary conditions for the velocity on the walls are used.
The tangential electrical currents and the radial component of the magnetic
field vanish on the conducting walls. For insulating walls the magnetic field must match the
external magnetic field.

\section{Hydrodynamics}

According to the  Rayleigh  condition the Taylor-Couette flow 
with the rotation law (\ref{basicOm}) is stable for 
\begin{equation}
\mu_\Omega>\hat\eta^2.
\label{ray}
\end{equation}
This condition has been extended to stratified fluids by Ooyama (1966).
The general condition for nonaxisymmetric solutions 
is not known. It is known, however,  that the axisymmetric mode is the most unstable
mode for dissipative Taylor-Couette flows for $\mu_\Omega>0$ and 
the nonaxisymmetric  modes can be  more unstable than the axisymmetric ones only for
counter-rotating cylinders (Langford et al. 1988).
 \begin{figure}[htb]
   \includegraphics[width=8.0cm, height=8.0cm]{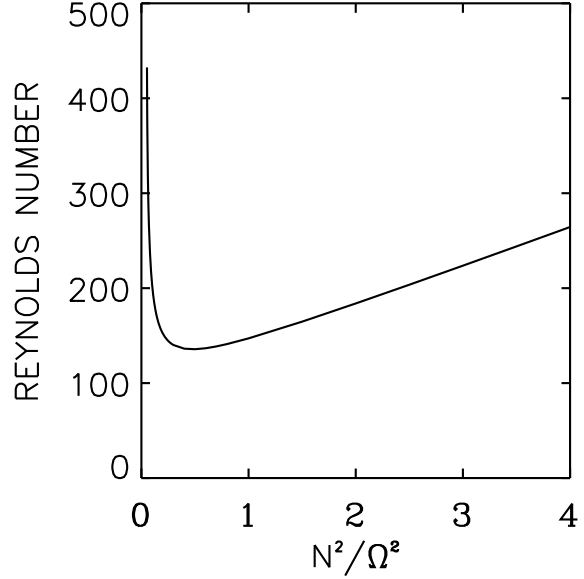}
    \caption{The marginal stability line for $m=1$ disturbances  in a  container
 with $\hat\eta=0.5$ and $\mu_\Omega=0.3$. The instability is  stabilized for both too weak  and too strong stratifications. The minimum Reynolds number belongs to a Froude number of 1.4  depending on the rotation law. }
    \label{NN}
 \end{figure}

Surprisingly, a stable vertical density stratification destabilizes the Taylor-Couette flow with decreasing angular velocity ($\mu_\Omega<1$) even beyond the Rayleigh line. This stratorotational instability (SRI) is nonaxisymmetric,  the most unstable mode is the mode with $m=1$. 
Shalybkov \& R\"udiger (2005) found that the instability condition is 
$\mu_\Omega<\hat\eta$ instead of $\mu_\Omega<1$ which has experimentally been confirmed
 by Le Bars \&  Le Gal (2007). This finding  based on a restricted number of   calculations. More detailed calculations are here reported  with  new results extending the previous conclusions.

In Fig. \ref{NN} the marginal stability line (the line which separates stable and unstable regions) is
calculated as function of  $N^2/\Om_{\rm in}^2$ for $m=1$ disturbances and for a standard 
container with $\hat\eta=0.5$ and with 
$\mu_\Omega=0.3$. The latter value  
lies  beyond the Rayleigh line where nonstratified flows are  stable.

We are looking for  the minimum critical Reynolds number which exists for a buoyancy frequency $N/\Om_{\rm in}\simeq0.71$, or -- what is the same --  ${\rm Fr}\simeq1.4$.  Figure~\ref{NN}  reveals the SRI as a delicate balance of buoyancy and rotation law. Both the frequencies must approximately be equal otherwise one of the two stabilities dominates the constellation. 

\begin{figure}[htb]
   \vbox{  
   \includegraphics[width=8.0cm, height=6.0cm]{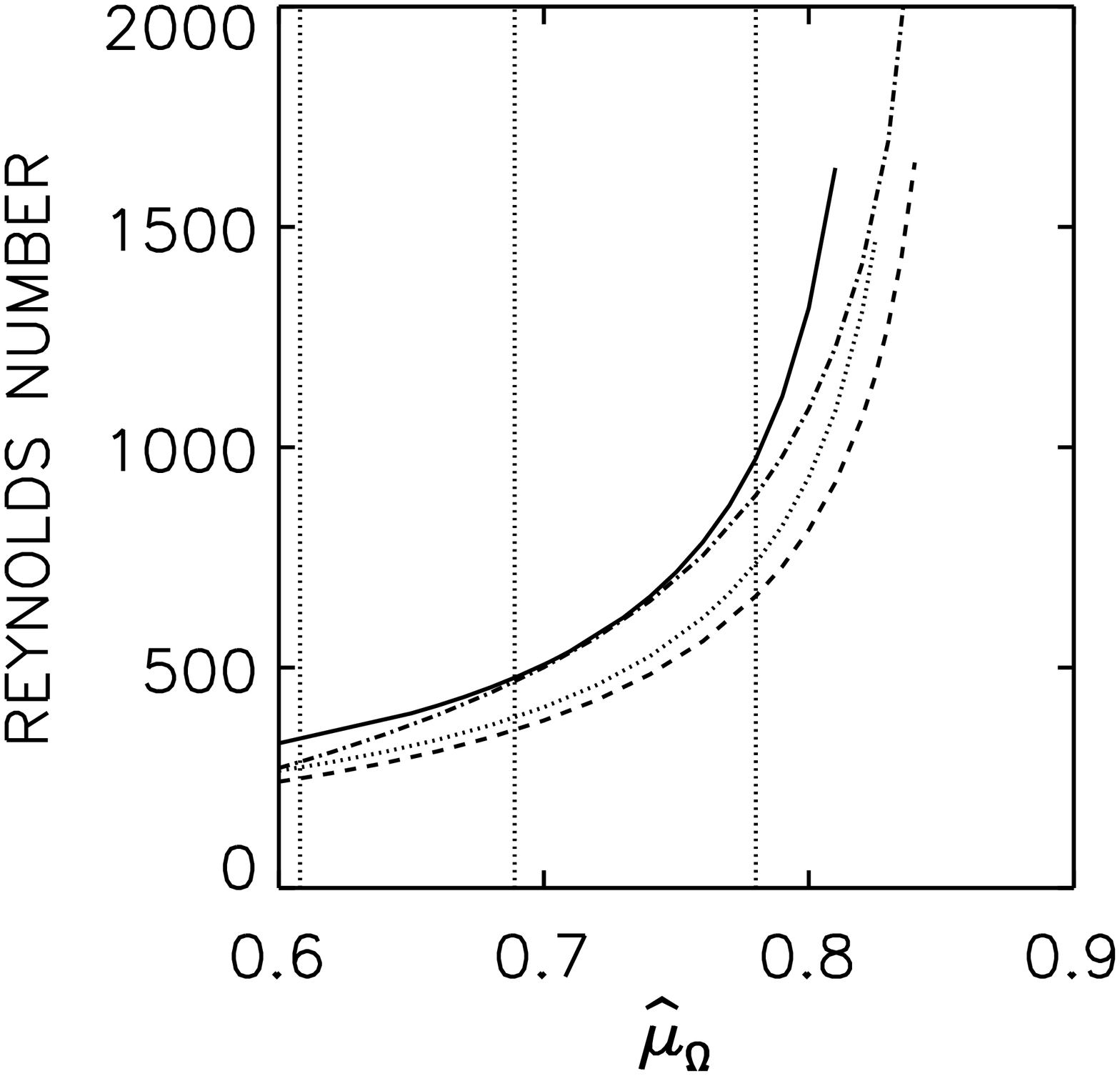}
    \includegraphics[width=8.0cm, height=6.0cm]{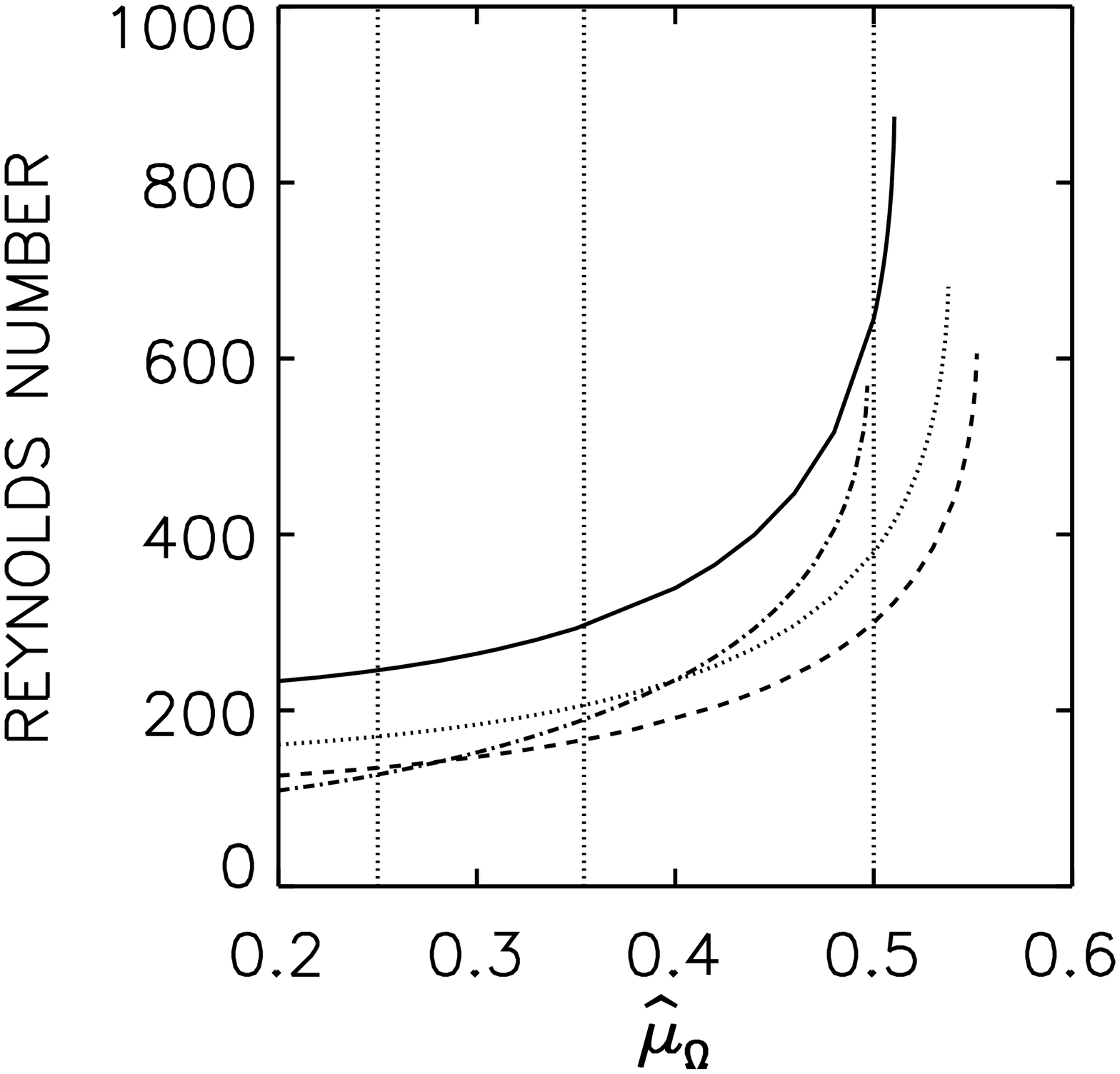}
    \includegraphics[width=8.0cm, height=6.0cm]{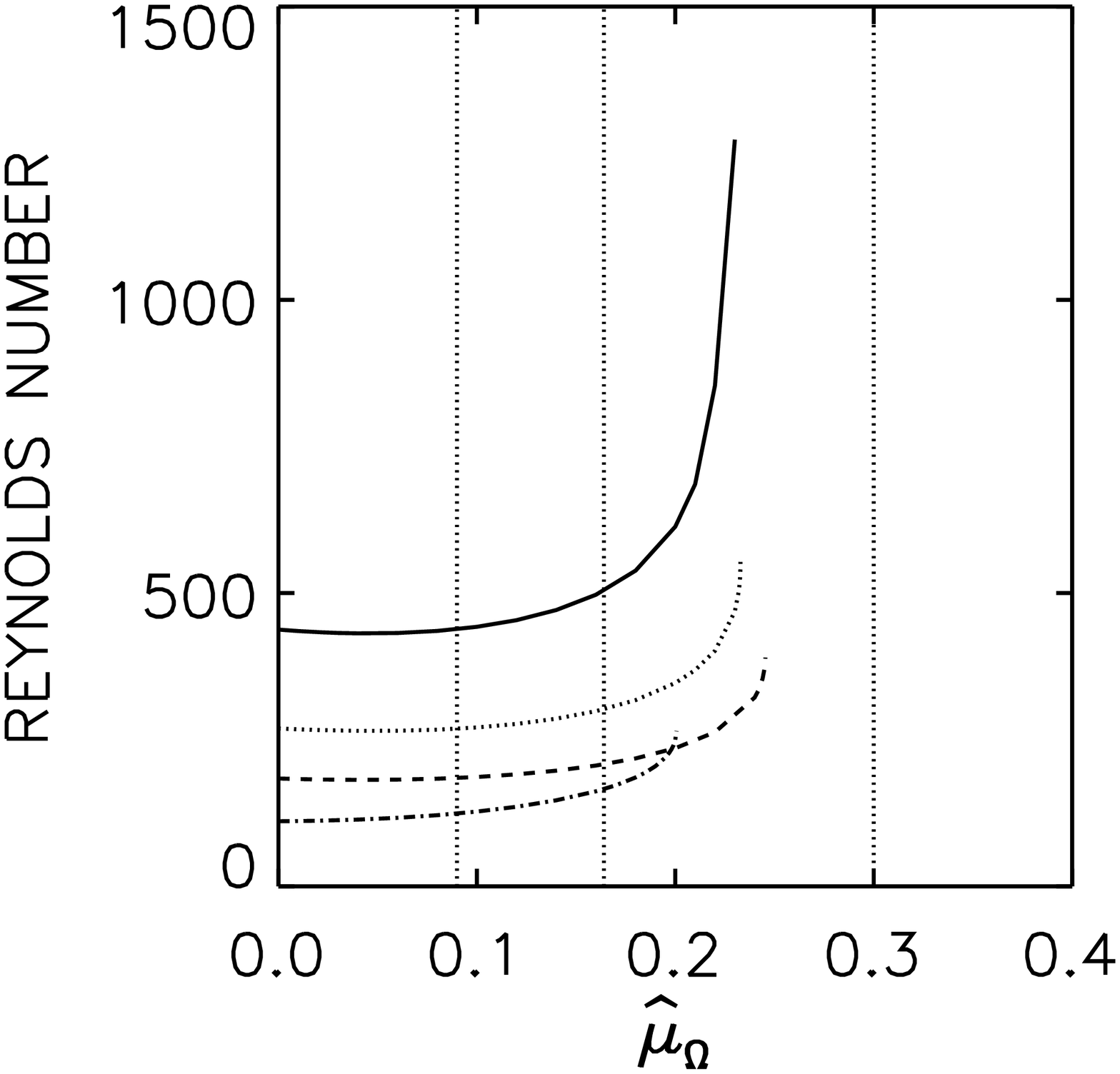}
    }
    \caption{The marginal stability lines for small-gap (top, $\hat\eta=0.78$), modest-gap ($\hat\eta=0.5$, middle) and  wide-gap ($\hat\eta=0.3$, bottom). The vertical dotted  lines denote the Rayleigh limit, the pseudo-Kepler rotation and the pseudo-galactic rotation ($\mu_\Omega=\hat\eta$).  The curves represent different density stratifications with $\rm Fr=2.2$ (dot-dashed),  $\rm Fr=1$ (dashed),  $\rm Fr=0.7$ (dotted),  $\rm Fr=0.5$ (solid, maximum stratification).  }
    \label{gaps}
 \end{figure}

The  question is how this  balance  depends  on the rotation law and on the geometry of the container. In  Fig. \ref{gaps}
 three containers are used with  different gap sizes. The computations also concern  various density stratifications. 
For any gap size three vertical  lines are  given. Left is the line for the Rayleigh limit where $\Om_{\rm out}= \Om_{\rm in} R_{\rm in}^2/R_{\rm out}^2$. The right line mimics the rotation law $\Om\propto 1/R$ which is typical for galaxies ($\Om_{\rm out}= \Om_{\rm in} R_{\rm in}/R_{\rm out}$). The central line represents the Kepler rotation law $\Om\propto R^{-1.5}$, i.e. $\mu_\Omega=\hat\eta^{1.5}$, which we shall call  the pseudo-Kepler limit. The main result of the Fig. \ref{gaps} is that
{\em the pseudo-Kepler rotation is always unstable}. This, however,  is not true for the more flat pseudo-galactic rotation which for large gaps becomes stable. From the bottom plot in Fig. \ref{gaps} we have to realize that the pseudo-galactic rotation law with $\Om\propto 1/R$ is already too flat for the possible existence of nonmagnetic SRI in  galaxies.
 
The present calculations do {\em not} confirm the $\hat\mu=\hat\eta$ line as the limit for the
SRI. This limit can be  larger for  small-gap containers and can be  smaller for
 wide-gap containers. The stability limit seems to approach the  Rayleigh line in the limit
of very wide gaps  and goes to unity for very
narrow gaps. We shall ask in the following whether such results are influenced by the existence of a toroidal magnetic field.

\section{SRI with  azimuthal magnetic field}
Without density stratification the ideal Taylor-Couette flow with imposed azimuthal magnetic field 
is stable against axisymmetric disturbances if  
\begin{equation}
\frac{1}{R^3}\frac{{\rm d}}{{\rm d}R}(R^2\Om)^2-\frac{R}{\mu_0\rho}\frac{\rm d}{{\rm d}R}\left(\frac{B_\phi}{R}
\right)^2>0
\label{mich}
\end{equation}
(Michael 1954). This condition  is the combination of the Rayleigh stability condition for differential rotation
without  magnetic field and the magnetic field stability condition without  rotation.
 Without a global rotation the magnetic field is stable (the second term in (\ref{mich}) is positive) for the magnetic field profile (\ref{basicB}) if 
\begin{equation}
0<\mu_B<\frac{1}{\hat\eta} \equiv \hat\mu_{0}.
\label{mu0}
\end{equation}
Therefore,  all magnetic profiles with positive $\mu_B$ or with $\mu_B<2$ (for $\hat\eta=0.5$) are stable against axisymmetric perturbations.
As the  dissipative effects are stabilizing  the flow
a magnetic field with $\mu_B$ beyond the interval (\ref{mu0})
becomes unstable against axisymmetric disturbances  in case that the  magnetic field amplitude
(or its Hartmann number) exceeds a  critical value.

The stability condition
for the azimuthal magnetic field  against nonaxisymmetric disturbances after  Tayler (1961) is
\begin{equation}
\frac{\rm d}{{\rm d}R}(RB_\phi^2)<0.
\label{tay}
\end{equation}
For the magnetic field profile (\ref{basicB}) it takes the form
\begin{equation}
0<\mu_B<\frac{4\hat\eta(1-\hat\eta^2)}{3-2\hat\eta^2-\hat\eta^4}
\equiv \hat\mu_{1}
\label{mu1}
\end{equation}
for the $m=1$ (kink) mode which is the most
unstable mode. Hence,  all azimuthal magnetic fields with positive $\mu_B$ smaller than 0.62 
(for $\hat\eta=0.5$) are stable against nonaxisymmetric disturbances. Current-free fields with $B_\phi\propto 1/R$, i.e. $\mu_B=0.5$ for $\hat\eta=0.5$
are thus always stable against $m=0$ and $m=1$.

As always  $\hat\mu_1<\hat\mu_0$  the stability interval (\ref{mu1}) is  smaller than
the stability interval (\ref{mu0}). In this sense the nonaxisymmetric (`kink') 
disturbances are more unstable than the axisymmetric (`sausage') disturbances. However, the situation is much more complex under the presence of (differential) rotation and which instability really dominates depends on the parameters of the problem (R\"udiger et al. 2007a,b). 

In the present paper  the influence of stable vertical density stratifications on the Taylor-Couette flow stability with imposed azimuthal magnetic fields is considered for three different
cases:
\begin{itemize} 
\item[1)] both the  rotation law  and the magnetic field are individually stable and both together are unstable  (i.e. $\mu_B=0.5$, no critical Hartmann number)
\item[2)] the magnetic field is stable for 
$m=0$  but unstable for $m=1$  (i.e. $\mu_B=1$, one critical Hartmann number)
\item[3)] the magnetic field is so steep  that it is unstable
for $m=0$ and for $m=1$ (i.e. $\mu_B=3$, two critical Hartmann numbers).    
\end{itemize}
  
\subsection{AMRI with density stratification}
The  azimuthal magnetic field with $B_\phi\propto 1/R$  (i.e. $\mu_B=0.5$ for $\hat\eta=0.5$) is considered  which is current-free between the cylinders  as the simplest  
example of an azimuthal magnetic field which is basically stable without  rotation for both axisymmetric and asymmetric
disturbances. 

\begin{figure}[h]
   \includegraphics[width=8.0cm, height=8.0cm]{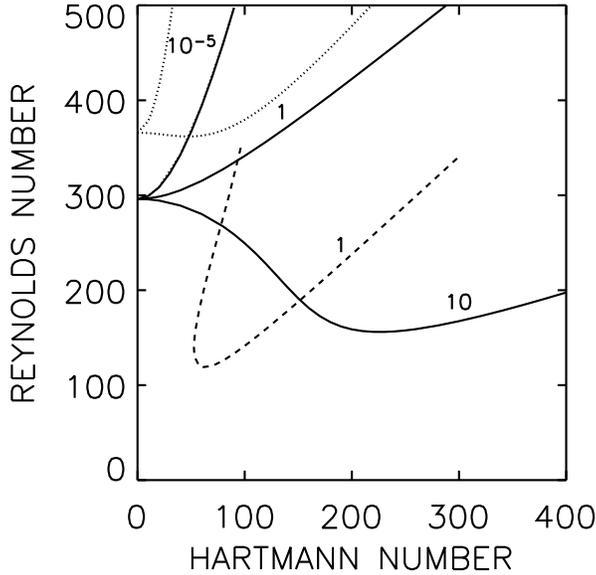}
    \caption{The marginal stability lines for the flow with {\em conducting} cylinders, $\hat\eta=0.5$,  current-free ($\mu_B=0.5$) azimuthal magnetic field,  pseudo-Kepler rotation ($\mu_\Omega=0.35$), $\rm Fr=0.5$ for disturbances with $m=1$ (solid) and $m=2$ (dotted). The curves are labeled by their magnetic Prandtl numbers. For comparison the dashed line demonstrates  AMRI ($m=1$) for  homogeneous fluids ($N=0$).}
    \label{amri}
 \end{figure}

\begin{figure}[h]
   \includegraphics[width=8.0cm, height=8.0cm]{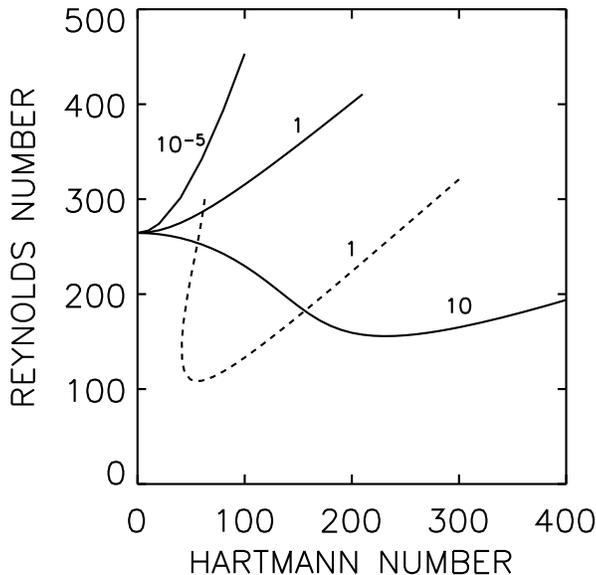}
    \caption{The same as Fig. \ref{amri} but for {\em insulating} cylinders with
$\mu_\Omega=0.3$ (as in Fig. \ref{NN}) but without $m=2$ mode.}
    \label{amri1}
 \end{figure}

The typical stability diagram is presented by the Figs. \ref{amri} and  \ref{amri1} for containers with
conducting/isolating cylinders, medium gap width ($\hat\eta=0.5$) and rather flat rotation law (pseudo-Kepler
rotation with  $\mu_\Omega=0.35$ and a slightly steeper rotation law with $\hat\mu=0.3$)
 which are stable without density stratification and without magnetic field. We find that the results depend only
slightly  on the conducting properties of the cylinder material.

The solid line in Figs.~\ref{amri} and \ref{amri1} is the marginal stability line of 
a homogeneous fluid with ${\rm Pm}=1$. There are no solutions at both the vertical axis and  the horizontal axis. The MHD  flow is only unstable as a combination of differential rotation and the 
azimuthal magnetic field. For any  supercritical Hartmann number  there are  two critical Reynolds numbers between which the fluid is unstable. Similarly, for any supercritical Reynolds number there are  two critical Hartmann numbers between which the fluid is unstable. We have called this phenomenon as the Azimuthal MagnetoRotational Instability  (AMRI) which scales with the magnetic Reynolds number in the same way as it is known from the standard MRI  in Taylor-Couette  experiments. However, magnetic Reynolds numbers of order 100 are too high to be realized in the MHD laboratory. 

With a density  stratification (${\rm Fr}=0.5$) the situation changes. The following  results can be interpreted as either AMRI for density-stratified fluids or as the influence of an azimuthal magnetic field on SRI.
There is now an instability
(SRI) without any magnetic field for ${\rm Re} > 296$ at $m=1$ and for ${\rm Re} > 366$ at $m=2$ (for conducting cylinders, pseudo-Kepler  rotation law, Fig. \ref{amri}) and for ${\rm Re} > 264$ at $m=1$ (insulating cylinders,  $\mu_\Omega=0.3$, Fig. \ref{amri1}). 
For this latter case the  calculations are restricted to  the kink ($m=1$) instability. 

\begin{figure}[h]
   \hbox{
   \includegraphics[width=4.5cm, height=4.5cm]{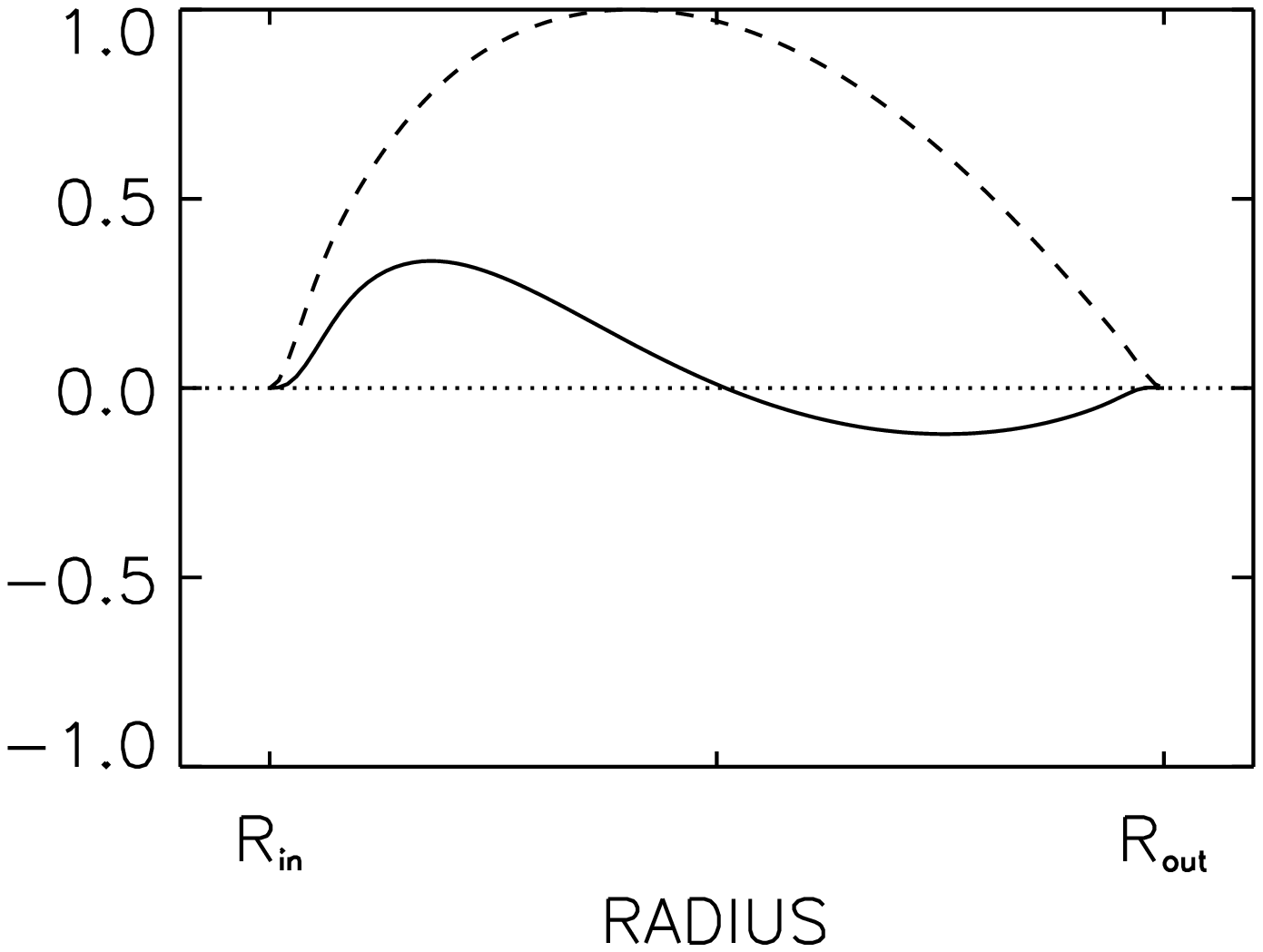}
   \includegraphics[width=4.5cm, height=4.5cm]{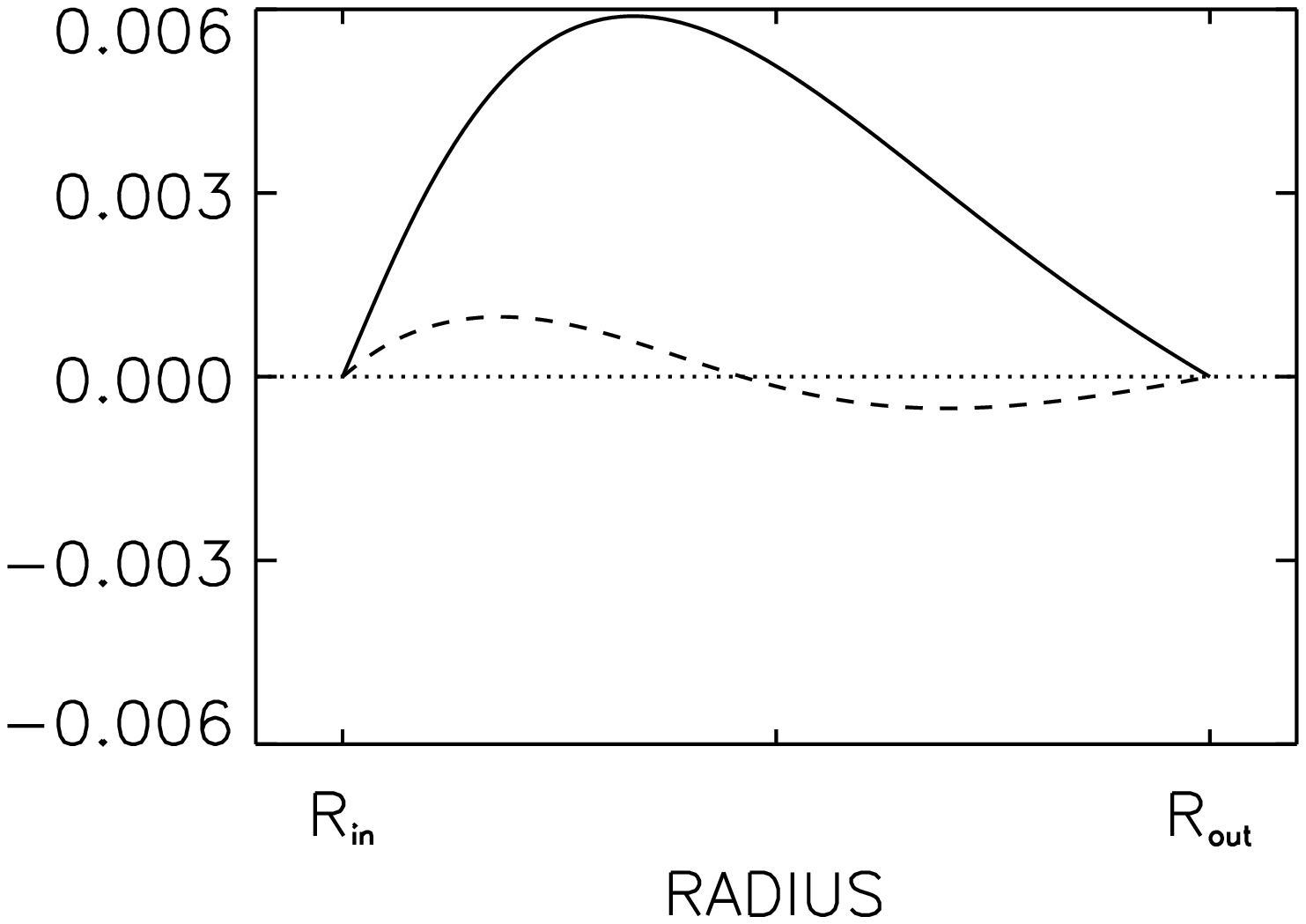}}
    \caption{The eigenfunctions for the radial components of flow (left) and field (right) for  $\rm Pm=1$. Solid: real part, dashed: imaginary part; the (arbitrary) amplitude is normalized. $\rm Fr=0.5$,  $\rm Ha=100$, $ m=1$, see Fig. \ref{amri}.}
    \label{profiles1}
 \end{figure}

\begin{figure}[h]
   \hbox{
   \includegraphics[width=4.0cm, height=4.0cm]{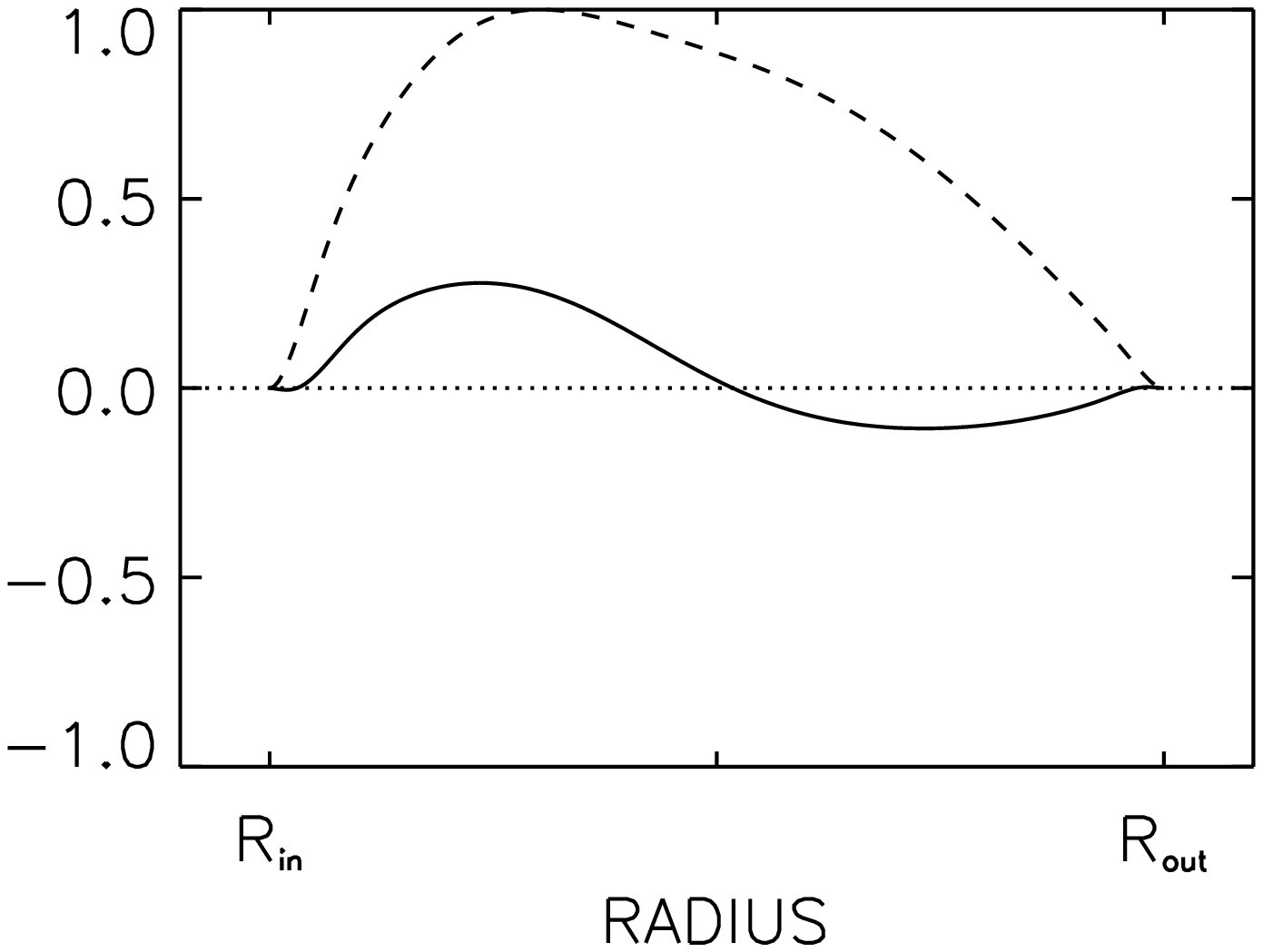}
   \includegraphics[width=4.0cm, height=4.0cm]{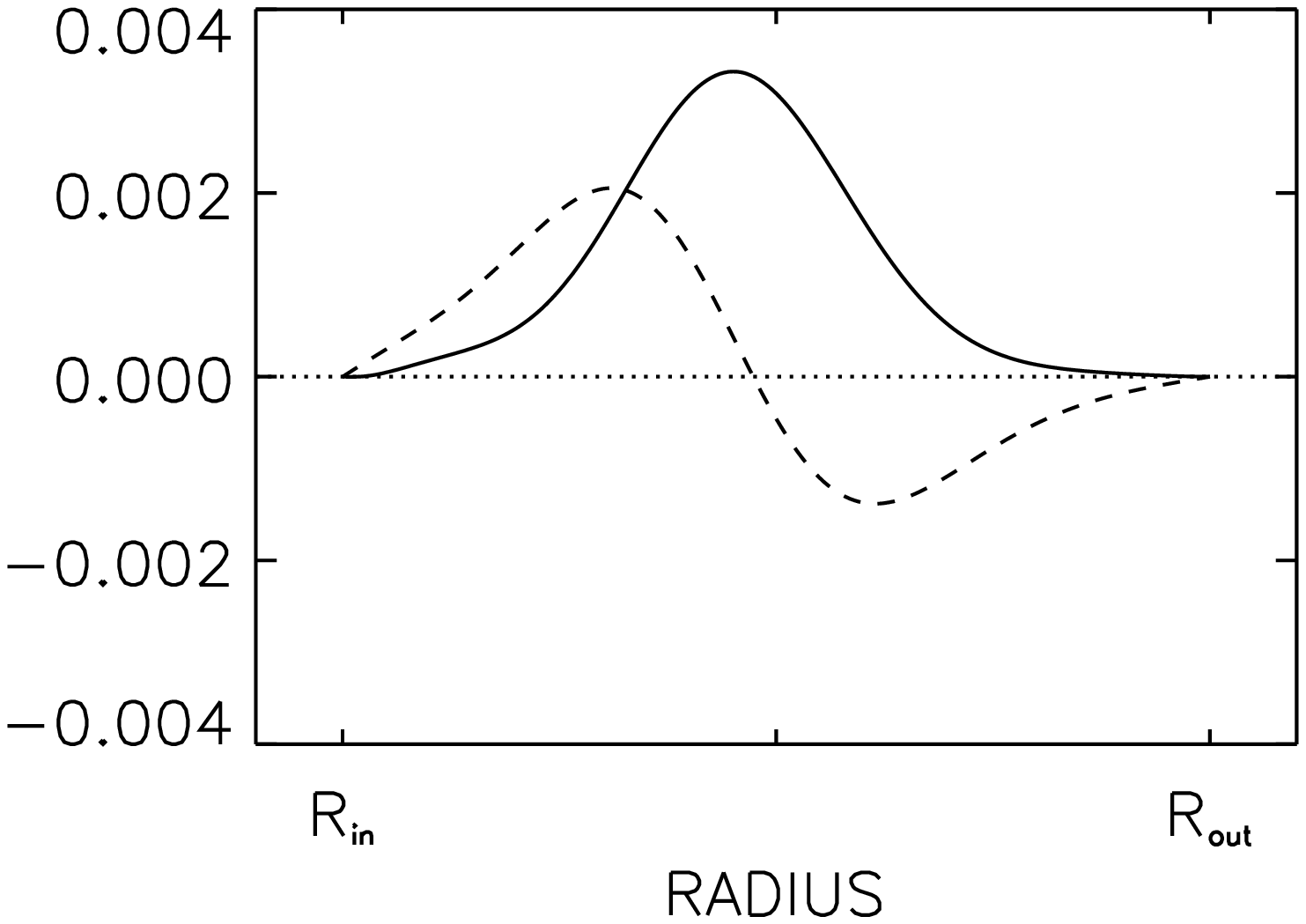}}
    \caption{The same as in Fig. \ref{profiles1} but for $\rm Pm=10$.  The eigenfunctions do not reflect a boundary phenomenon.}
    \label{profiles10}
 \end{figure}

For conducting boundaries (see Fig. \ref{amri}) and for a special case ($\rm Ha=100$, $m=1$) the radial eigenfunctions are given for   
 $\rm Pm=1$ (Fig. \ref{profiles1}) and $\rm Pm=10$ (Fig. \ref{profiles10}) which demonstrate that the SRI modes do not form  boundary phenomena. The profiles are normalized since in the linear theory the amplitudes have no own physical meaning. For the  linear growth time in units of the rotation period of the inner cylinder for the solution with $\rm Pm=1$ one obtains 
\begin{equation}
\frac{\tau_{\rm growth}}{\tau_{\rm rot}}\simeq \frac{243}{{\rm Re}-341}.
\label{growth}
\end{equation} 
Hence, for $\rm Re=400$ the e-folding time is  about four rotations. The vertical wave number at the critical Reynolds number $\rm Re=341$ is 8.63 what means that the vertical cell size in units of the gap width is about 0.36. Without magnetic field the wave number is 8.16. Common action of magnetic field and density stratification leads to rather flat cell configurations. 


Note that the influence of the azimuthal magnetic field on the SRI strongly depends on the magnetic
Prandtl number Pm. For ${\rm Pm}\leq 1$ the SRI is stabilized by the magnetic field
but it can be {\em supported}  for  ${\rm Pm}>1$ (see Figs.~\ref{amri}, \ref{amri1}). For ${\rm Pm}=10$ the critical Reynolds number is reduced  (by a factor of  two) for ${\rm Ha}\simeq 200$. In this case AMRI supports the SRI. This behavior is similar to  that of the standard MRI
with axial magnetic field and  hydrodynamically unstable rotation where also for ${\rm Pm}>1$ the magnetic field supports the centrifugal instability (R\"udiger \&  Shalybkov 2002).

Note also that the magnetic  suppression of the SRI for small Pm is rather weak. Even for ${\rm Pm}=10^{-5}$ a (large) Hartmann number of order of 100 only leads to a  small  critical Reynolds number of about 500. The lines for ${\rm Pm}=10^{-5}$ in Figs.~\ref{amri} and  \ref{amri1} are identical with the lines for all smaller  Pm, i.e. they  also hold  for  gallium  ($10^{-6}$) and also   mercury  ($10^{-7}$). Here all the curves scale  with Re  rather than Rm so  that the corresponding magnetic Reynolds numbers are quite small compared with those for AMRI. As also the corresponding Hartmann numbers with $\rm Ha\lsim 100$ are not too high, laboratory experiments with fluid metals should  be possible provided a sufficiently density stratification can be produced. 

It is also shown in Fig. \ref{amri} that for a given Pm the critical Reynolds numbers for $m=2$ exceed  those for $m=1$. The differences, however, are not too big so that for slightly supercritical Re several modes should  be excited. This is insofar surprising as the known smoothing action of differential rotation upon  modes with high $m$ here seems to be very weak.

\begin{figure}[htb]
   \includegraphics[width=8.0cm, height=8.0cm]{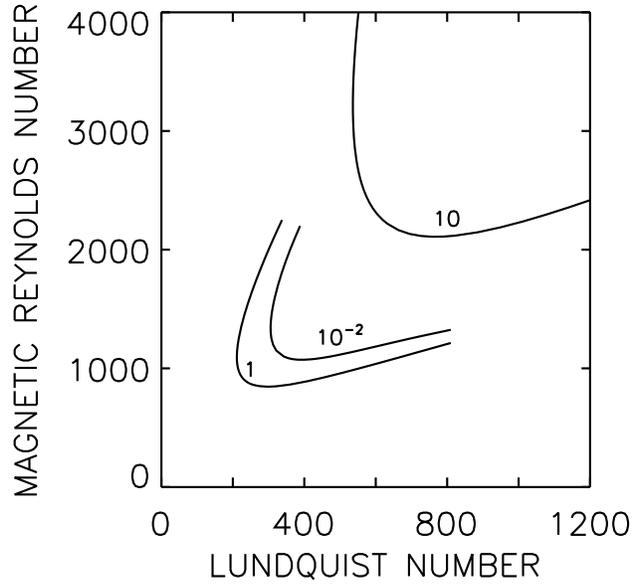}
    \caption{AMRI for stratified fluids with ${\rm Fr}=0.5$ between {\em insulating} cylinders
($\hat\eta=0.5$,  $\mu_B=0.5$). The rotation law is so flat ($\mu_\Omega=0.6$) that nonmagnetic SRI does not exist (see Fig. \ref{gaps}, middle). Lines are labeled by their Pm numbers.
}
    \label{rml}
 \end{figure}

The rather small Reynolds numbers and Hartmann numbers  lead to the  impression that the instability should  be observable into the laboratory. For hydrodynamically unstable situations we have
only a suppression of the instability by the magnetic field for small Pm. For hydrodynamically stable situations (solid lines in Figs. \ref{amri} and
\ref{amri1}) it was shown  for uniform fluids  that the appropriate numbers of the problem  are the magnetic
Reynolds number Rm and the Lundquist number S
\beg
{\rm Rm}={\rm Pm}\cdot{\rm Re}, \q {\rm S}={\rm Pm}^{1/2} \cdot {\rm Ha},
\label{rm}
\ende 
rather than Re and Ha. Figure \ref{rml} demonstrates that the same is true for density-stratified  fluids with so flat   rotation laws  that hydromagnetic SRI does not operate (see Fig. \ref{gaps}). Both the critical magnetic Reynolds number and Lundquist number do not vary remarkably when the magnetic Prandtl number varies over 3 orders of magnitude.

 Stratified flows without magnetic field
become stable for   $\mu_\Omega$-values  slightly greater than $\mu_\Omega=0.5$. 
According to Fig. \ref{rml} the flow is stable for $\mu_\Omega=0.6$  without a magnetic field but it becomes unstable with magnetic fields.  The critical Rm are not monotonous with Pm.  The values of Rm decrease with Pm for small Pm and v.v. The result can be seen in Fig. \ref{rml} which differs in this respect  to the AMRI without density stratification (see R\"udiger et al. 2007a).

\begin{figure}[htb]
   \includegraphics[width=8.0cm, height=8.0cm]{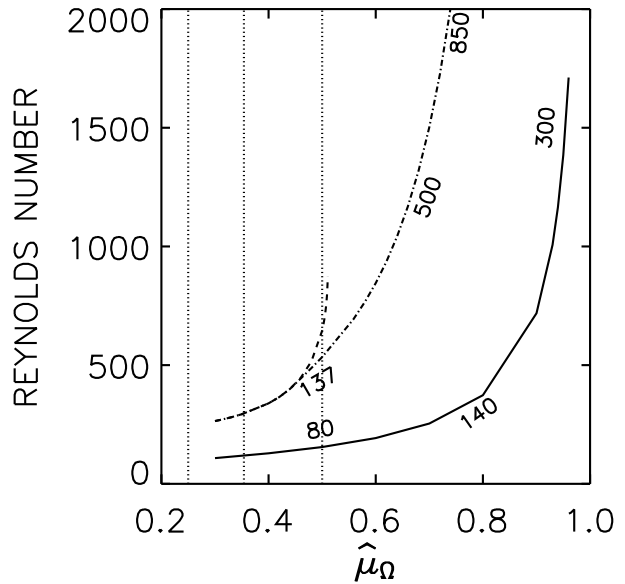}
    \caption{The minimum Reynolds numbers  for { insulating} cylinders,  $\hat\eta=0.5$, ${\rm Pm}=1$  and for the kink ($m=1$) mode.
The vertical lines are the same as in Fig. \ref{gaps}. The numbers on the curves are Hartmann numbers which correspond to the minimum Reynolds numbers. The solid line is AMRI without a stratification, the  dot-dashed line is AMRI with stratification  (${\rm Fr}=0.5$, see Fig. \ref{rml}) and the dashed line is SRI without  magnetic field. }
    \label{field}
 \end{figure}

The solid line in Fig. \ref{field} demonstrates
that the AMRI (like the standard MRI) exists for all rotation laws with decreasing angular velocity as a function of radius (i.e. for $\mu_\Omega<1$).

 According to the Figs. \ref{amri} and  \ref{amri1} the marginal stability lines have always a minimum for
some Hartmann number including ${\rm Ha}=0$. These minimum Reynolds numbers are plotted in Fig.
\ref{field} as functions of $\mu_\Omega$. The dashed line is SRI without
magnetic field which disappears for too flat rotation laws ($\mu_\Omega\gsim 0.52$).
There is a smooth transition, however, to the instability AMRI (current-free toroidal field plus differential rotation) which  has been computed here for the first time with a density gradient (dot-dashed). The comparison with standard AMRI without ${\rm d}\rho/{\rm d}z$ (solid line) yields  the expected suppression  by density stratification. Note, however, that for quasi-galactic rotation ($\mu_\Omega=0.5$ in Fig. \ref{field}) the SRI {\em with} magnetic field can easier be excited than  without magnetic field. As  the figure shows the necessary magnetic field has a Hartmann number of about 137.

As AMRI scales with Rm rather than Re.   the  `smooth' transition interval  from the hydrodynamic solutions  to the magnetohydrodynamic   scales with  Pm$^{-1}$. Hence, the dot-dashed line in Fig. \ref{field} becomes more and more steep for decreasing Pm. 
\subsection{Density stratification and Tayler instability ($m=1$)}

\begin{figure}[htb]
\vbox{
   \includegraphics[width=8.0cm, height=8.0cm]{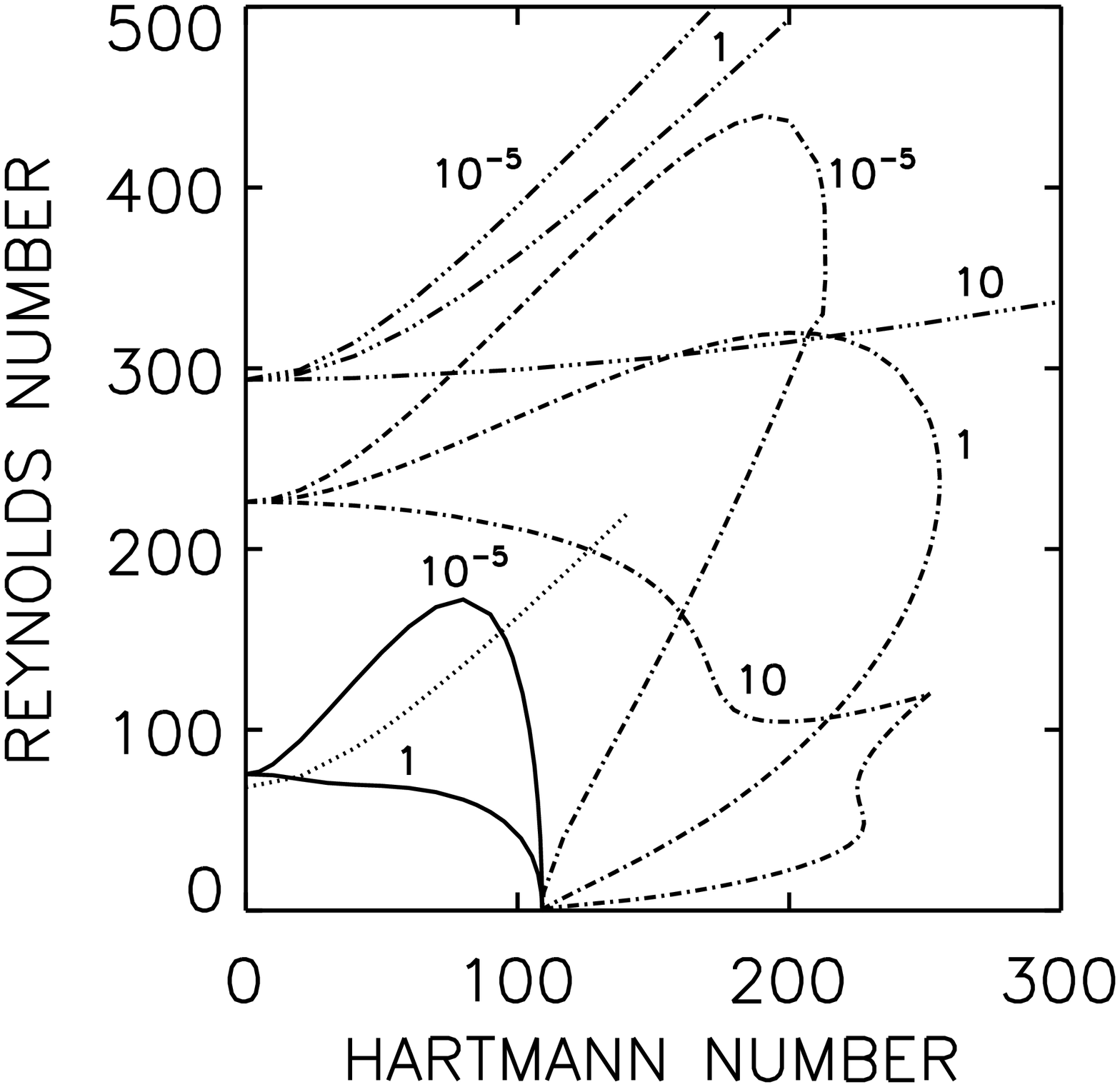}
    \includegraphics[width=8.0cm, height=8.0cm]{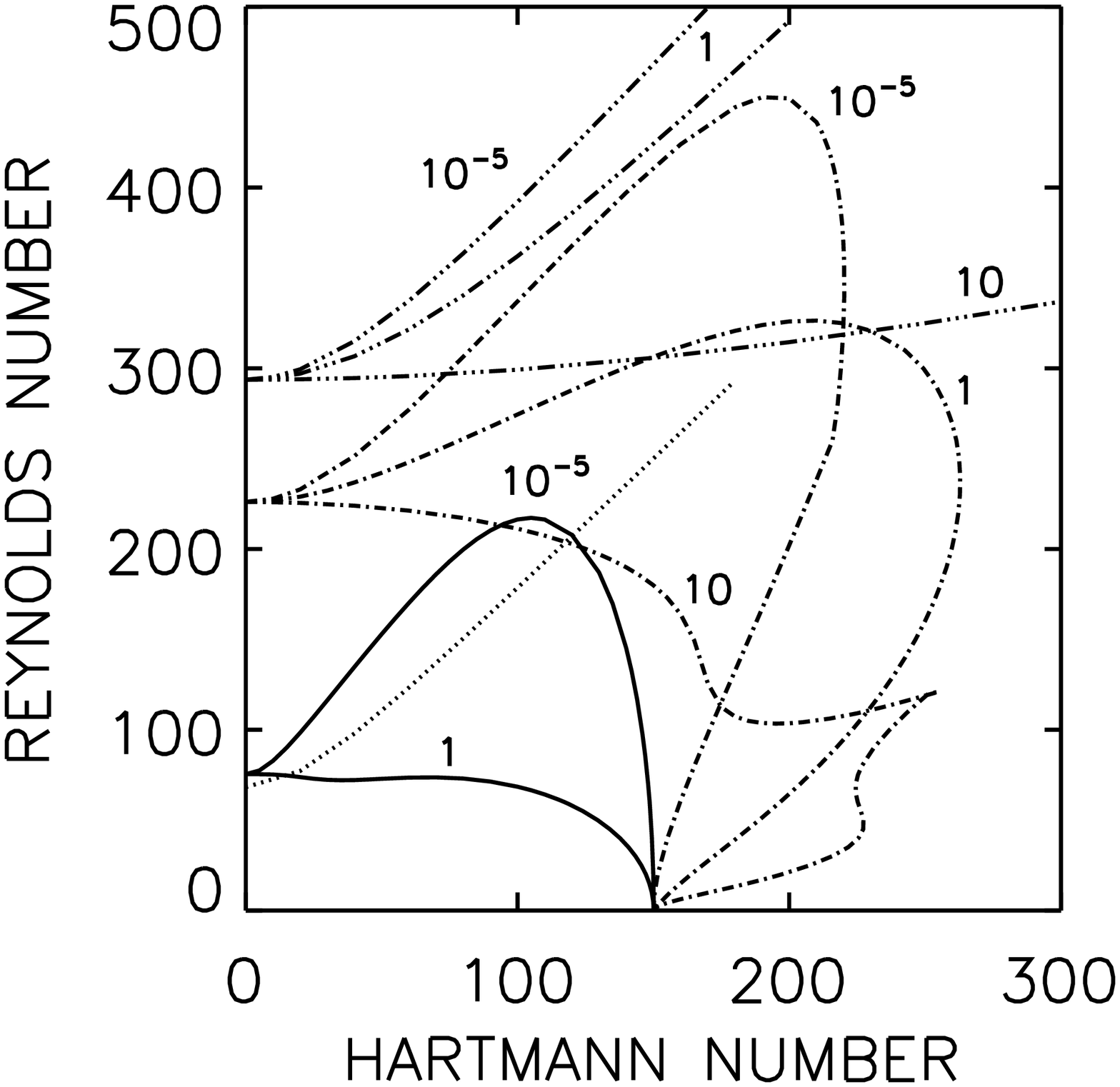}}
    \caption{The marginal stability lines for a flow with insulating  (top) and conducting (bottom) cylinders. It is  $\hat\eta=0.5$, $\mu_\Omega=0$ (resting outer cylinder)  and $\mu_B=1$. The curves are labeled by their magnetic Prandtl  Pm. The $m=0$ (dotted) mode (no dependence on Pm) and $m=1$ (solid) mode for homogeneous fluid and $m=0$ (dot-dot-dot-dashed) and $m=1$ (dot-dashed) modes for density-stratified fluids with ${\rm Fr}=0.5$.}
    \label{taync}
 \end{figure}

The influence of the vertical density stratification on the nonaxisymmetric 
Tayler instability (i.e. $\hat\mu_1<\mu_B<\hat\mu_0$) is considered for  rotation laws which are steep enough to be Rayleigh unstable.
Again the  gap width is $\hat\eta=0.5$, the outer cylinder is resting ($\mu_\Omega=0$) and the toroidal field is rather homogeneous 
($\mu_B=1$) so that without rotation only the mode $m=1$ is unstable.
Calculations have been made for both insulating  and conducting cylinders
(Fig. \ref{taync}). The difference between conducting
and insulating boundary conditions proves to be only  quantitative.
 
For $m=1$  the  critical Hartmann numbers above which the flow is unstable for $\rm Re=0$ are 150 for conducting cylinders and 109 for insulating cylinders. These critical  numbers are not influenced by the density stratification.  For $m=0$ no critical Hartmann numbers here
  exist;  the  magnetic field stabilizes this Rayleigh-mode for both  
homogeneous  (dotted line) and density-stratified (dot-dot-dot-dashed lines)  fluids. For homogeneous fluids the (dotted) marginal stability line 
for axisymmetric disturbances does not depend on the magnetic Prandtl number Pm (Shalybkov 2006)!

The critical Reynolds numbers for the Rayleigh instability  for $\rm  Ha=0$  do not depend on the conducting properties of the cylinders.  They are  68 ($m=0$) and 75 ($m=1$)  without  stratification. 
The density-stratification {\em stabilizes}  the flow; the critical Reynolds numbers  are  294  for $m=0$
and 226  for $m=1$   (for ${\rm Fr}=0.5$).  
With density-stratification    the  kink mode ($m=1$) proves to be  the most unstable mode.

Moreover, for given Pm the kink mode is the most unstable one for all Ha for ${\rm Fr}=0.5$. For small Pm this mode is stabilized by  weak magnetic fields  before
a dramatic  destabilization happens for larger magnetic fields. For  increasing Pm  the  flow becomes unstable for decreasing Reynolds numbers at a fixed
Hartmann number. The opposite is also true: the smaller the Pm the stronger is the magnetic field influence
and the  flow becomes unstable for smaller Hartmann numbers at a fixed Reynolds number. 

Note that the competition between magnetic and centrifugal instabilities can lead to a rather
complex behavior as illustrated by the marginal stability line with ${\rm Pm}=10$. Generally, however,
the stability region is much larger for density-stratified fluids so that we find   that a stable
stratification increases the flow stability. Of particular interest is the phenomenon that for density-stratified fluids generally the critical Hartmann numbers are larger than without rotation and the same is true for small Pm with respect to the critical Reynolds number  (`ballooning'). For large magnetic Prandtl number, however, the magnetic field basically {\em destabilizes} the Rayleigh instability. Even for weak fields the critical Reynolds number is smaller than the Reynolds number for the nonmagnetic case.

\subsection{Density stratification and Tayler instability ($m=0,1$)}
\begin{figure}[htb]
   \vbox{  
   \includegraphics[width=8.0cm, height=6.0cm]{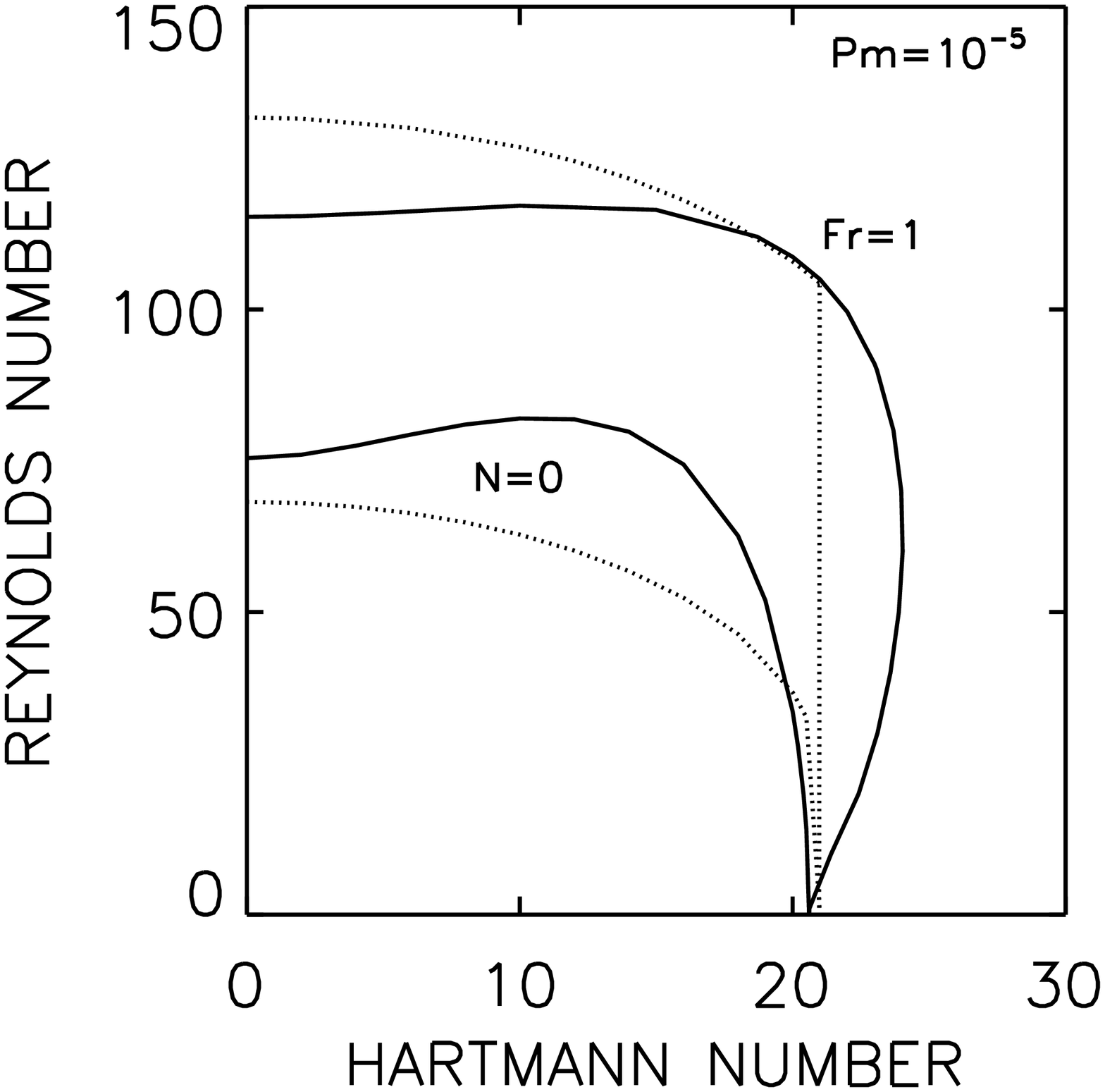}
    \includegraphics[width=8.0cm, height=6.0cm]{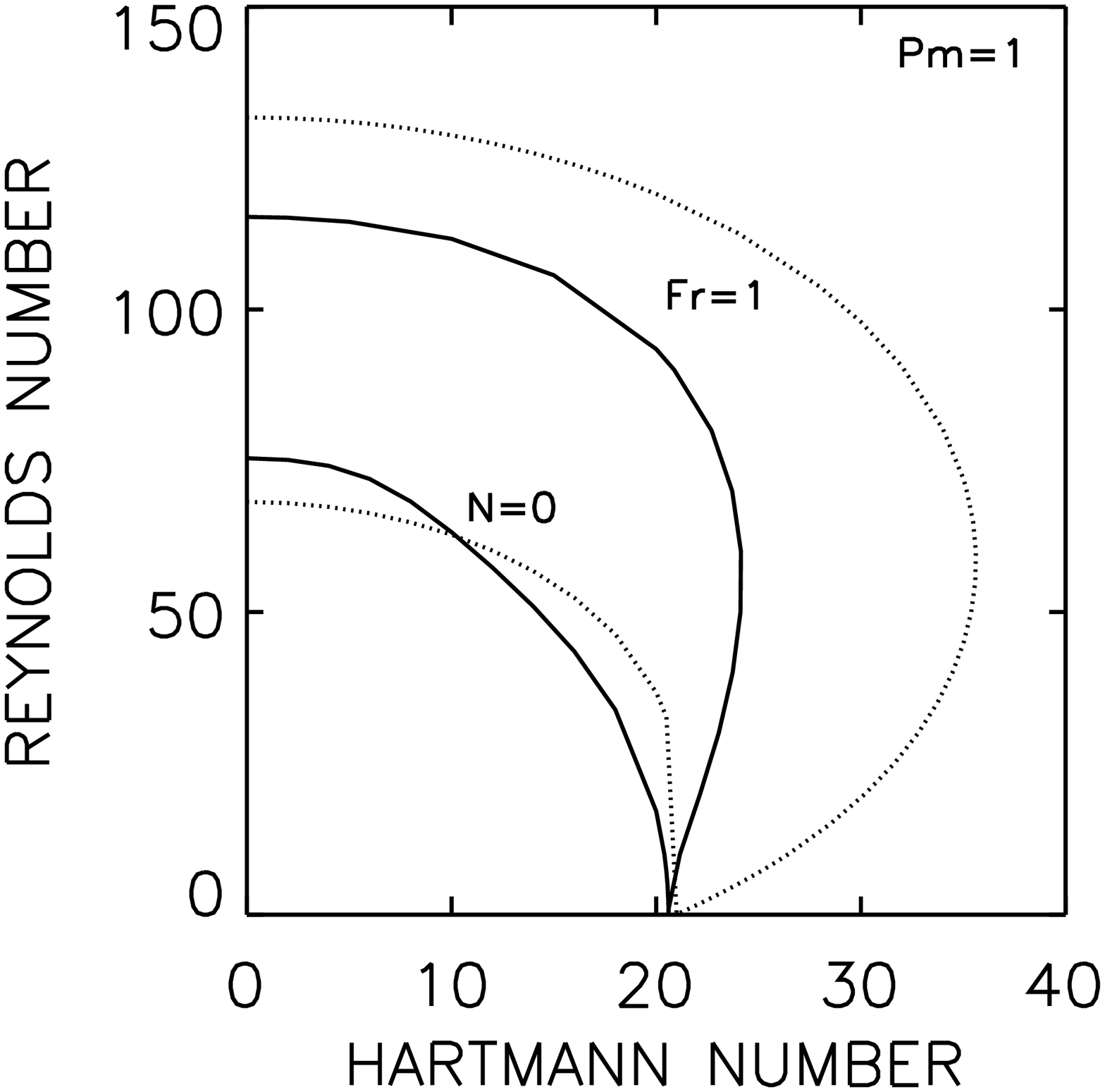}
    }
    \caption{The marginal stability lines for {\em conducting} cylinders with $\hat\eta=0.5$,
$\mu_\Omega=0$, $\mu_B=3$, ${\rm Pm}=10^{-5}$ (top) and ${\rm Pm}=1$ (bottom). The lines connect the Rayleigh instability without magnetic field with the Tayler instability without rotation. 
The dotted lines are for $m=0$; the solid lines   for $m=1$. The inner curves are for homogeneous fluids ($N=0$) and the outer curves  for stratified fluids with $\rm Fr=1$. The density stratification stabilizes in both directions.
}
    \label{gen}
 \end{figure}

The most complex situation appears when   both the modes with  $m=0$  and $m=1$ are Tayler-unstable. Only conducting cylinders with medium-sized 
gap ($\hat\eta=0.5$), outer cylinder at rest  and a toroidal  magnetic field
with strong currents ($\mu_B=3$) are  analyzed. The differential rotation is Rayleigh unstable without magnetic field and density stratification  and the magnetic field is Tayler-unstable without rotation (for $m=0$ and $m=1$).  The critical Hartmann number for $\rm Re=0 $ is nearly equal for all cases: $m=0$ gives ${\rm Ha}_0=21$ and  $m=1$ 
gives ${\rm Ha}_1=20.6$. For other parameters it can be more different. Either the $m=0$ or the $m=1$ mode can be the most unstable mode. Often also the $m=0$ mode is the most unstable
for small Ha numbers and fast differential rotation (see R\"udiger et al.  2007b).

The critical Reynolds numbers above which the rotation becomes
unstable without a magnetic field ($\rm Ha=0$) are called  Re$_0$ and 
Re$_1$. Re$_0$ is smaller than Re$_1$ for flows with 
$0<\mu_\Omega<\hat\eta^2$.
The density stratification  increases Re$_0$
and Re$_1$ (Fig. \ref{gen}). Note that with density-stratification  the $m=1$ mode becomes
the most unstable mode. With magnetic field it is also the most unstable mode almost
everywhere except for ${\rm Pm}=10^{-5}$ and ${\rm Ha}>{\rm Ha}_0$.

Note that  `ballooning' of the stability region is produced by the  density stratification.
The  density stratification in combination with the basic rotation stabilizes the flow for larger  magnetic fields. Under the influence of differential 
rotation and density stratification stronger magnetic field amplitudes prove to be stable than they were without rotation. The effect already exists for rather slow rotation rates (see Fig. \ref{gen}, $\rm Pm=1$). This phenomenon is  a consequence of the explicit inclusion  of the stable density stratification into the MHD equations.  One finds that both the Rayleigh instability and the Tayler instability are stabilized by the density stratification. Without rotation, however, the density influence of the Tayler instability would remain negligibly small.

\section{Electromotive force for AMRI}
Stratorotational instability under the influence of a toroidal field is tempting to apply the concept of the mean-field  electrodynamics in turbulent fields. The nonaxisymmetric components of  flow and  field can be used as fluctuations while the axisymmetric components are considered as the mean quantities. Simply the averaging procedure is the integration over the azimuth $\phi$. 

It is standard to express the turbulence-induced electromotive force as
\begin{equation}
\vec{\cal E}= \langle\vec{u'} \times \vec{B'}  \rangle = \alpha \circ \langle\vec{B}\rangle - \eta_{\rm T}{\rm curl} \langle\vec{B}\rangle
\label{emf1}
\end{equation}
with the alpha-tensor $\alpha$ and the (scalar) eddy diffusivity $\eta_{\rm T}$. In  cylindric geometry the mean magnetic field $\langle\vec{B}\rangle $ has only a $\phi$-component and the mean current ${\rm curl}\langle\vec{B}\rangle$ only has  a $z$-component. Hence, ${\cal E}_\phi= \alpha_{\phi\phi}  \langle{B}_\phi\rangle$ and ${\cal E}_z=
- \eta_{\rm T}{\rm curl}_z \langle\vec{B}\rangle=0$. The latter  relation only holds for AMRI where the mean magnetic field is current-free.
\begin{figure}[htb]
   \vbox{  
   \includegraphics[width=9.0cm, height=7.0cm]{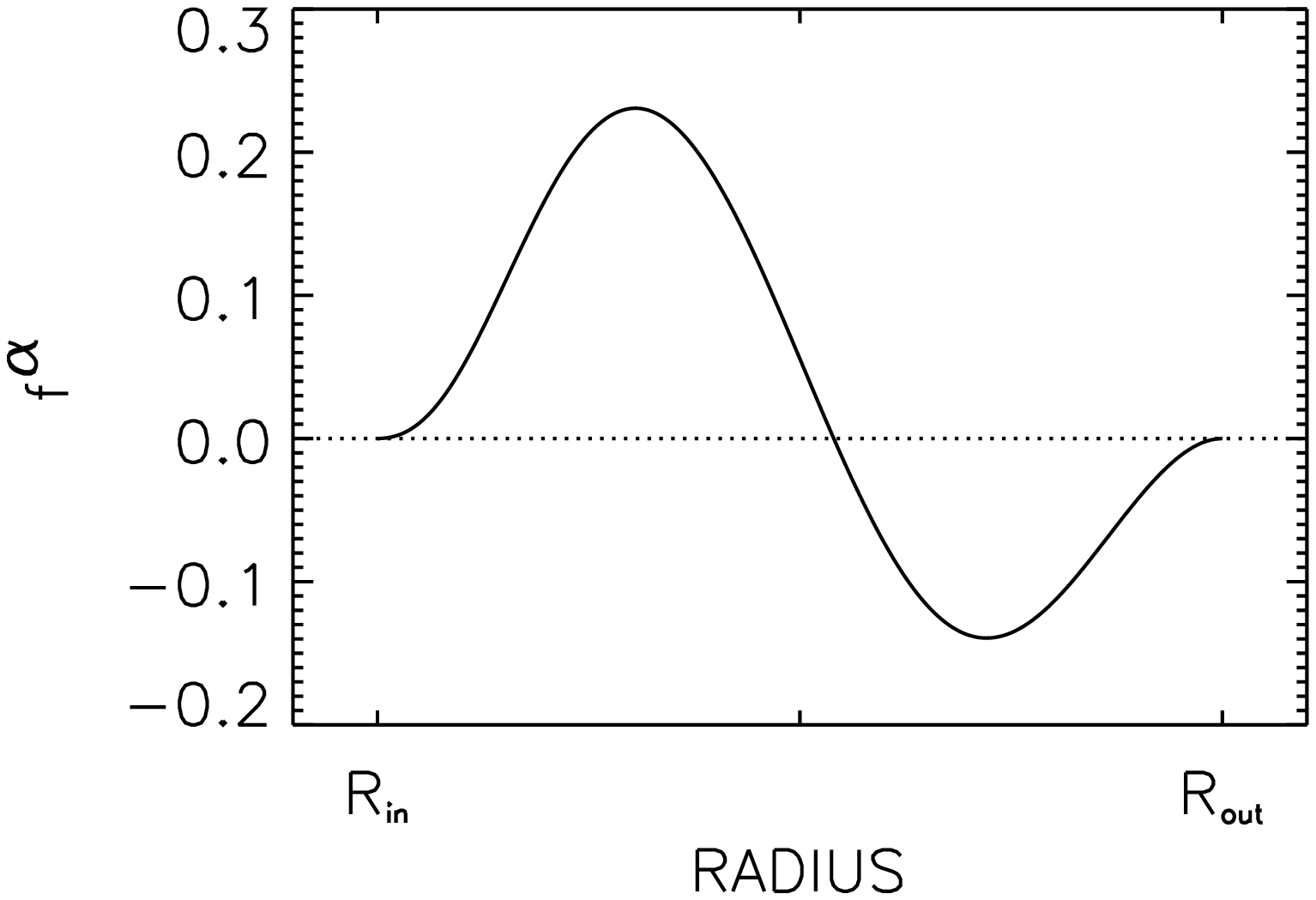}
    \includegraphics[width=9.0cm, height=7.0cm]{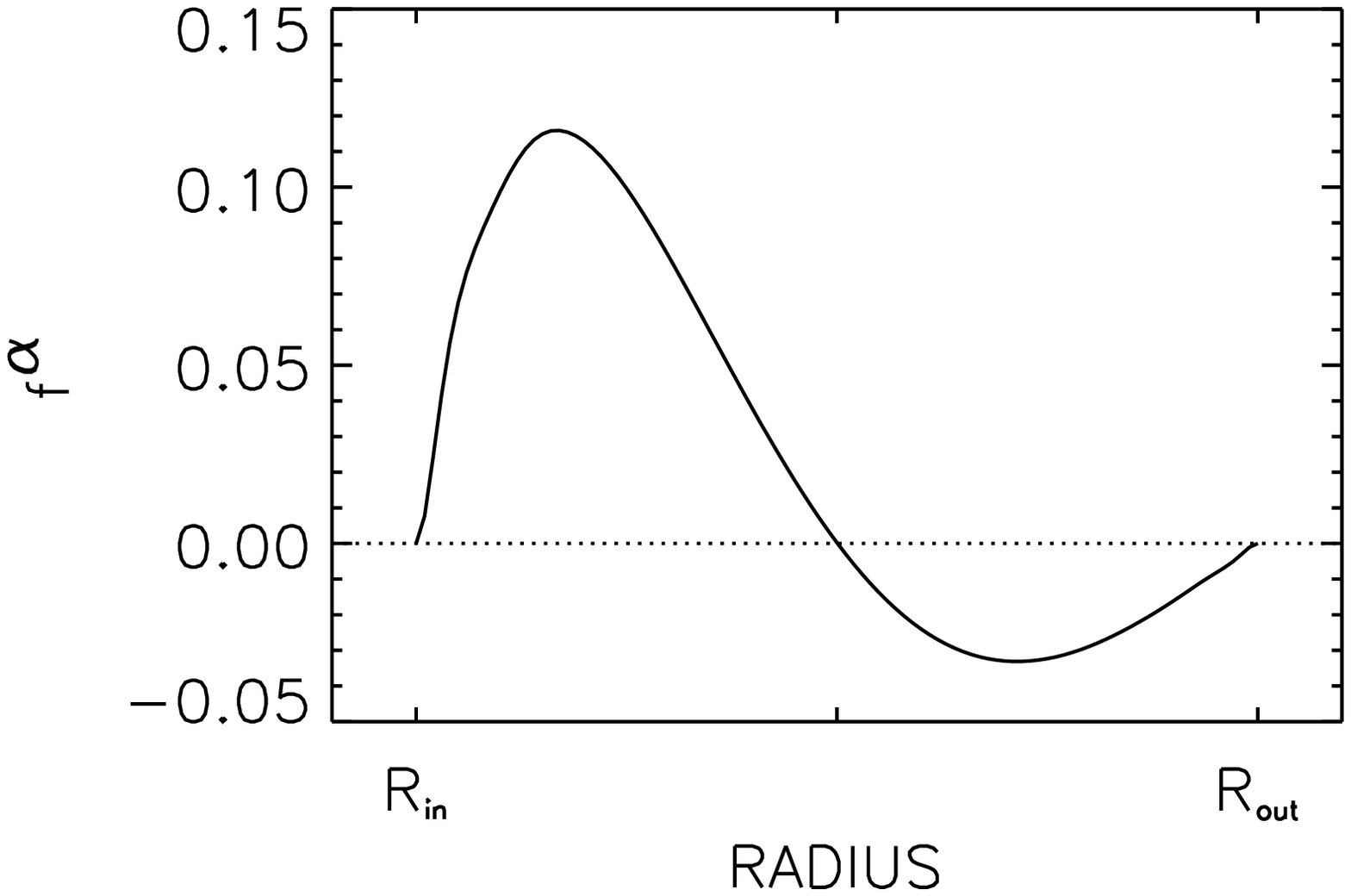}
    }
    \caption{The correlation function (\ref{emf2}) for AMRI with { conducting} cylinders and with $\hat\eta=0.5$,
$\mu_\Omega=0.35$ (pseudo-Kepler), $\mu_B=0.5$, ${\rm Pm}=1$. Density stratification is zero (top) and finite (bottom, $\rm Fr=0.5$). The Reynolds numbers $\rm Re=141$ (top) and  $\rm Re=341$ (bottom) were taken from Fig. \ref{amri}.
}
    \label{emf}
 \end{figure}

In the present paper we only consider the alpha-effect  in the frame of a linear theory where all functions are free to one and the same (complex)  arbitrary parameter.   This is only possible if the second-order quantities such as  ${\cal E}_\phi$ are normalized with a second-order  quantity. In the considered case it makes sense to form the correlation function
\begin{equation}
{ f}^\alpha= \frac{\langle\vec{u'} \times \vec{B'}  \rangle_\phi}{ {\rm MAX}(\sqrt{\langle \vec{u}'^2 \rangle\langle \vec{B}'^2 \rangle   }) },
\label{emf2}
\end{equation}
where all the correlations along the radius $R$ are normalized with one and the same parameter. Hence,  this function does no longer contain the arbitrary factor of the eigenfunctions  and  by definition it is smaller than unity. The main questions are the sign of this quantity and the influence of the density stratification. Note that the  $B_\phi$ has been  given as positive in the model  setup where rotation axis and density gradient are antiparallel. The latter is quite characteristic for the situation at the poles of rotating stars or disks so that  the old question whether an alpha-effect at the poles does exist or not is here concerned.

One can easily show that second-order correlations of quantities running with ${\rm exp}({\rm i}(kz+m\phi))$ after integration over $\phi$ do not depend on the coordinate $z$. The term $kz $ only fixes the phase where the integration starts.

The correlation function $ f^\alpha$ can be estimated for fast-rotating magnetoconvection in the  high conductivity limit $\alpha\simeq \sqrt{\langle \vec{u}'^2 \rangle}$ and with $\langle \vec{B}'^2 \rangle \simeq (\eta_{\rm T}/\eta) \langle \vec{B} \rangle^2$ after Krause \& R\"adler (1980)   
so that 
\begin{equation}
{ f}^\alpha <\sqrt{\frac{\eta}{\eta_{\rm T}}}.
\label{emf3}
\end{equation}
For stellar convection this is  a small number. Hence, the numerical constraints for the exact calculations of the eigenfunctions are very high. One must keep this finding in mind when discussing the numerical results given in Fig. \ref{emf} for an example without density stratification and another one  with density stratification. For simplicity the magnetic Prandtl number is put to unity. Note the  influence of the  density stratification. The examples are taken from the Fig. \ref{amri} for $\rm Ha=100$. The resulting correlations for $N=0$ are much more antisymmetric  than the values for $\rm Fr=0.5$. Averaged over the radius the $ f^\alpha$ vanishes for $N=0$ but not for the presence of a density stratification. 

This cannot be a boundary 
effect. A boundary effect is  implausible as the preferred radial direction is perpendicular to the rotation axis for cylinders which mimics the equator in rotating spheres and there the $\alpha$-effect {\em must} vanish.  Indeed, the antisymmetry is reduced if a density stratification is allowed (Fig. \ref{emf}, bottom). Then  the density gradient is now  the preferred direction in the system  parallel to the rotation  allowing the formation of large-scale helicity (as in rotating spheres at the poles). We are thus tempted to assume that the presented calculations for the presence of a density gradient indeed have demonstrated for the existence of an $\alpha$-effect for the magnetic SRI.

\begin{figure}[htb]
   \includegraphics[width=9.0cm, height=7.0cm]{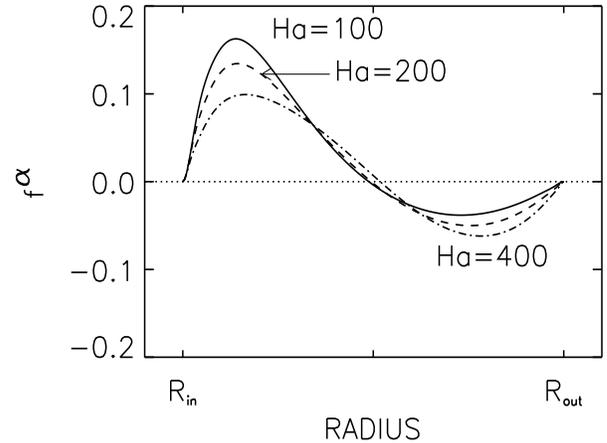}
    \caption{The correlation function (\ref{emf2}) for  $\hat\eta=0.5$,
$\mu_\Omega=0.35$ (pseudo-Kepler), $\mu_B=0.5$, ${\rm Pm}=0.01$. $\rm Fr=0.5$). The Reynolds numbers are $\rm Re=531$ for $\rm Ha=100$  and  $\rm Re=999$ for  $\rm Ha=200$  and  $\rm Re=2137$ for  $\rm Ha=400$. Note that the $alpha$-effect is quenched for stronger magnetic fields.
}
    \label{emfpm}
 \end{figure}

Figure \ref{emfpm} presents similar results but for the small magnetic Prandtl number $\rm Pm=0.01$. The correlation functions do provide a dominance of the positive contributions but for higher and higher magnetic fields the difference becomes smaller. It seems that  the numerical results  do here reflect a magnetic quenching of the $\alpha$=-effect.

\section{Conclusions}
The stability of the dissipative Taylor-Couette flow under the joint influence of
a stable vertical density stratification and an azimuthal magnetic field is considered. The problem is of interest for future laboratory experiments but also within the frame of  accretion disk physics. The Kepler rotation generates strong toroidal magnetic fields   dominating the poloidal components. The standard MRI which works with only axial fields may be of minor relevance for the stability of the Kepler rotation law compared with the azimuthal MRI in connection with the density stratification and Tayler instability. Mainly nonaxisymmetric `kink' modes ($m=1$) are considered but there are also examples where the axisymmetric modes   are the most unstable ones.

We started with a    discussion of  the SRI  without magnetic fields. 
For a flat rotation law 
a  stratification value Fr exists for  which the critical Reynolds number of the rotation has 
a minimum (see Fig. \ref{NN}). The SRI is basically stabilized by both 
too weak or  too strong stratification. The instability only appears if the characteristic buoyancy time approaches the rotation period.
Also if the rotation is too flat the instability disappears.

The limiting ratio $\mu_\Omega$  strongly depends on the gap width
(see Fig. \ref{gaps}).  For small gaps  rotation laws with $\mu_\Omega>\hat\eta$ even prove to be unstable while for wide  gaps the condition  $\mu_\Omega<\hat\eta$ results for exciting SRI.
Our previous finding that $\mu_\Omega \lsim \hat\eta$ limits the 
   SRI is reproduced for medium-sized gaps. In all cases, however, the quasi-Kepler rotation proves to be unstable.  New experiments with different gap sizes and larger
Reynolds numbers could  verify these results.

 Figures~\ref{amri} and \ref{amri1} yield the basic results for SRI subject to toroidal fields. The magnetic field is assumed as current-free in the fluid between the cylinders, i.e. $B_\phi \propto 1/R$ excluding  Tayler instability. Without density  stratification  no Rayleigh instability exists for quasi-Kepler rotation but nonaxisymmetric modes with $m=1$ are unstable as a result of the interaction of differential rotation and magnetic field (`AMRI').  
 
The magnetic influence strongly depends on the magnetic Prandtl number. The instability needs higher Reynolds number  for ${\rm Pm}\lsim 1$ and it needs lower Reynolds number  for ${\rm Pm}\gsim 1$. After  our experiences with MHD instabilities this is not a surprise. It is a surprise, however, that always the magnetic influence is only weak.  Up to Hartmann numbers of ${\rm Ha}\simeq 100$ only a magnetic-induced factor of two plays a role\footnote{${\rm Ha}=100$ for gallium corresponds to $B\approx 2200/R_0\ [{\rm Gauss}]$ with $R_0$ in cm}. Hence, our conclusion is that the SRI  survives for rather high magnetic fields. For large Pm it is even supported by the toroidal magnetic field.

The combination of rotation, density stratification
and magnetic field leads to complex results.  However, a basic observation is that 
 the critical Reynolds numbers, if  existing, above which the flow becomes unstable
without a magnetic field, are increased by the stable density stratification.
In contrast, the critical Hartmann numbers, if  existing by the Tayler instability, above which the field
becomes unstable without a rotation, do not depend on the stratification. Different transition are possible between the two limits.

Generally speaking, the  density stratification `balloons' the stability region and in this sense
it stabilizes the flow. For slow rotation  the maximal stable  magnetic field exceeds the critical magnetic field without rotation while for faster rotation the   maximal stable  magnetic field is smaller than this critical value.
Even rather slow values of the Reynolds numbers lead to a stabilization of those fields which are unstable for $\rm Re=0$. The effect is strong for $\rm Pm=1$ but it becomes smaller for decreasing magnetic Prandtl numbers. 

For steep radial profiles of the magnetic field (i.e. strong axial currents) and magnetic Prandtl numbers $\rm Pm \gsim 1$ one 
also finds  the magnetic field destabilizing the Rayleigh instability, i.e. the critical  Reynolds numbers with magnetic field are lower than the Reynolds numbers without magnetic field.  This magnetic destabilization only exists for not too small magnetic Prandtl numbers. It  exists for both uniform and density-stratified fluids (Fig. \ref{gen}). 

Finally, the $\phi$-component of the electromotive force representing  the $\alpha$-effect of the mean-field dynamo theory has been computed. The computations require an extreme degree of accuracy. The results demonstrate the importance of the density stratification for the existence of the $\alpha$-effect. Without density stratification the correlations are vanishing in the radial average. With an axial density stratification  the calculations model the polar region of a rotating sphere or disk. This interpretation accepted we found the  $\alpha_{\phi\phi}$ at the northern pole as {\em positive} (Fig. \ref{emf}). Again, the basic ingredient of this $\alpha$-effect model is the density stratification.


\begin{thebibliography}{}
\bibitem{C96}
Curry, C., \& Pudritz, R.E. 1996, MNRAS, 281, 119
\bibitem{Dub05}
Dubrulle, B., Mari\'e, L., Normand, Ch., Richard, D., Hersant, F., \& Zahn, J.-P. 2005, A\&A, 429, 1
\bibitem{}
Krause, F., \& R\"adler, K.-H. 1980,
Mean-field magnetohydrodynamics and dynamo theory
(Pergamon Press, Oxford)
\bibitem{L88}
Langford, W.,   Tagg, R.,   Kostelich, E.,   Swinney, H., \& Golubitsky, M. 1988,  Phys. Fluids,
 31, 776 
\bibitem{LL07}
Le Bars, M., \&  Le Gal, P. 2007, Phys. Rev. Lett.,  99, 064502
\bibitem{M54}
Michael, D. 1954, Mathematika,  1, 45
\bibitem{MMY01}
Molemaker, M.J.,  McWilliams, J.C.,  \& Yavneh, I.  2001, Phys. Rev. Lett.,  86, 5270 
\bibitem[1996]{OP96}
Ogilvie, G.I., \&  Pringle, J.E. 1996, MNRAS,  279, 152 
\bibitem[1966]{O66}
Ooyama, K.J. 1966,  J. Atmos. Sci.,  23, 43 
\bibitem{PT97}
Papaloizou, J.C.B., \& Terquem, C. 1997, MNRAS, 287, 771
\bibitem[1985]{PT}
Pitts, E., \& Tayler, R.J. 1985, MNRAS, 216, 139
\bibitem[2002]{RS02}
R\"udiger, G., \&  Shalybkov, D. 2002, Phys. Rev. E,  66, 016307 
\bibitem[2007a]{REA07}
R\"udiger, G.,   Hollerbach, R.,  Schultz, M., \&  Elstner, D. 2007a, MNRAS,
    377, 1481 

\bibitem[2007b]{REA07b}
R\"udiger, G.,  Schultz, M.,  Shalybkov, D., \& Hollerbach, R. 2007b, Phys. Rev. E,
  76, 056309 

\bibitem[2006]{S06}
Shalybkov, D. 2006, Phys. Rev. E,  73, 016302
\bibitem[2005]{SR05}
Shalybkov, D., \&  R\"udiger, G. 2005, A\&A,  438, 411  

\bibitem[1961]{T61}
Tayler, R.J. 1961, J. Nucl. Energ. C,  3, 266 
\bibitem{U06}
Umurhan, O.M. 2006,  MNRAS, 365, 85
\bibitem[2001]{YMM01}
Yavneh, I.,  McWilliams, J.C., \&  Molemaker, M.J.  2001, JFM,  448, 1 


\end{thebibliography}
\end{document}